# Multi-Track Time-Series Burst-Overlap Interferometry for Resolving Horizontal Deformation in Earthquake-Cycle Studies

**Xing Li[1], Zhuang Gao[2], Han Chen[3], Zhangfeng Ma[4], Yangkang Chen[1], Alexandros Savvaidis[1]**

[1] Bureau of Economic Geology, Jackson School of Geosciences, University of Texas at Austin, 78758, Austin, United States

[2] Department of Earth and Space Sciences, Southern University of Science and Technology, Shenzhen, China, 518055

[3] State Key Laboratory of Earthquake Dynamics and Forecasting, Institute of Geology, China Earthquake Administration, Beijing, China, 100029

[4] Earth Observatory of Singapore, Nanyang Technological University, Singapore, 639798

Corresponding author: Zhangfeng Ma (zhangfeng.ma@ntu.edu.sg)

**Key Points:**

- A novel Multi-track Time-Series Burst-Overlap Interferometry (MTSB) framework is evaluated across earthquake-cycle applications.
- MTSB products referenced to ITRF2014 show centimeter-level GNSS agreement for displacement and millimeter-per-year agreement for velocity.
- Across different earthquake cycle studies, MTSB improves constraints on regional plate kinematics and fault kinematics.

**Abstract**

Interferometric Synthetic Aperture Radar (InSAR) is intrinsically insensitive to the north-south component of crustal motion because of its near-polar geometry, which makes accurate quantification of this deformation a persistent limitation of InSAR. We introduce a Multi-track Time-Series Burst-Overlap Interferometry (MTSB) framework that delivers absolute, ITRF-referenced deformation fields across representative interseismic, coseismic, and postseismic applications by combining optimized phase linking, a unified time-series framework for separating deformation and residual misregistration, and block-wise spectral analysis. Case studies spanning co-, inter-, and postseismic phases demonstrate that MTSB accurately resolves along-track horizontal deformation, with primary sensitivity to north–south motion, while substantially reducing orbit-related artifacts. Independent comparisons with GNSS indicate centimeter-level agreement for postseismic and coseismic displacement estimates and millimeter-per-year agreement for interseismic velocities. The resulting deformation fields improve constraints on regional plate kinematics and fault kinematics, facilitate more reliable Euler-pole estimation, and provide by-product orbital corrections for conventional InSAR processing.

**Plain Language Summary**

Earthquakes move the ground in different directions, but ordinary satellite radar is much less sensitive to north-south motion than to east-west or vertical motion. This missing component limits our ability to measure how faults accumulate strain before earthquakes, how the ground moves during rupture, and how the crust continues to deform afterward. We developed a new approach, Multi-track Time-Series Burst-Overlap Interferometry (MTSB), to better measure horizontal ground motion by combining observations from several Sentinel-1 satellite paths, improving radar-signal quality, and correcting small orbit-related errors. The resulting measurements are tied to a global reference frame, enabling consistent comparison with ground-based GNSS observations and other geodetic datasets. By providing more complete deformation maps throughout the earthquake cycle, MTSB can improve constraints on fault-based seismic-hazard models, which inform long-term risk reduction, infrastructure resilience planning, and emergency preparedness in earthquake-prone regions.

## 1 Introduction

Geodetic observations of the earthquake cycle, encompassing co-, inter-, and postseismic phases, are critical not only for seismic-hazard assessment but also for understanding fault behavior, including strain accumulation, fault coupling, and postseismic stress transfer. Interferometric Synthetic Aperture Radar (InSAR) has become a key tool, delivering high-resolution, wide-area maps of surface displacement *(Avouac, 2015; Bürgmann et al., 2000; Elliott et al., 2016; Ryder et al., 2007)*. Yet, the near-polar orbits of radar satellites restrict conventional InSAR's ability to resolve the north-south component of ground motion. To overcome this limitation, three SAR-based approaches are widely employed to retrieve horizontal displacements along the satellite's azimuth (along-track) direction: azimuth pixel-offset tracking, Multi-Aperture Interferometry (MAI), and Burst-Overlap Interferometry (BOI) *(Bechor and Zebker, 2006; Grandin et al., 2016; Hu et al., 2008; Jung et al., 2010; Michel and Rignot, 1999)*. Each exploits different characteristics of SAR data to capture the elusive north-south component of ground motion, yet they differ markedly in accuracy, sensitivity, and practical applicability.

Azimuth pixel-offset tracking estimates offsets by maximizing cross-correlation between two SAR amplitude image patches *(Gray et al., 2001)*. It is widely used for mapping large surface displacements, particularly coseismic deformation, landslides, and glacier motion, because it can accommodate large displacement gradients without requiring interferometric phase continuity. Its accuracy, about one-tenth of the SAR pixel size *(Simons and Rosen, 2007)*, yields roughly 10-20 cm precision in the high-resolution Sentinel-1 data, which is sufficient for large coseismic shifts but inadequate for detecting subtle interseismic motions along slow faults. MAI creates forward- and backward-looking images through azimuth sub-aperture processing, yet its limited azimuth bandwidth of about 50 Hz for ScanSAR mode and ~320 Hz for TOPS mode *(Wegmüller et al., 2009)* restricts it to capturing only large along-track displacements. Although MAI has successfully mapped coseismic displacements of major earthquakes *(Barbot et al., 2008; Erten et al., 2010)*, its typical precision, ranging from several centimeters to the decimeter level, makes it unsuitable for interseismic studies. By contrast, BOI derives along-track deformation by double differencing consecutive burst overlaps and has demonstrated millimeter-level capability to resolve horizontal motions along slow-deforming faults *(Li et al., 2021)*. Together, these methods illustrate a trade-off between measurement range and precision in SAR-based along-track deformation observations. Pixel-offset tracking and MAI are generally well suited to large coseismic or surface-rupture displacements, whereas BOI offers higher precision for subtle deformation but requires careful treatment of coherence loss, misregistration, and reference-frame biases (Table S1).

The key principle is that Sentinel-1 TOPS mode illuminates each target within the burst overlap region with two full beam patterns whose spectral frequencies differ by roughly 5 kHz, an order of magnitude larger than typical sub-aperture separations, providing exceptional sensitivity to azimuth misregistration. This property underpins the Enhanced Spectral Diversity (ESD) technique *(Scheiber and Moreira, 2000)*, which refines azimuth offsets using the interferometric phase during fine image coregistration. By averaging a constant value from all burst-overlap interferograms, ESD effectively removes errors caused by orbit inaccuracies or ionospheric delay and has become the standard method for azimuth misregistration correction. However, accurate azimuth displacement can only be retrieved by averaging the phase bias across burst overlaps under the assumption that these regions are either deformation-free or exhibiting deformation below the measurement noise *(Prats-Iraola et al., 2012)*. In tectonically active areas with significant ground motion, this assumption breaks down, and the necessary averaging introduces fundamental biases. As a result, ESD suppresses genuine tectonic signals such as plate motion or rapid fault slip *(De Zan et al., 2014; Ma et al., 2022; Prats-Iraola et al., 2012)*. For example, along a fault slipping at 2 cm/yr over three years, the resulting azimuth misregistration of ~0.0043 pixels is removed during fine coregistration, rendering the true deformation undetectable.

This suppression is especially critical for large-scale tectonic processes, where deformation gradients extend hundreds of kilometers. A further limitation emerges when combining multiple satellite orbits, a common requirement for continent-scale deformation studies. Although along-track measurements are nominally tied to the ITRF2014 reference frame, each track estimates burst-overlap misregistration independently, effectively operating within its own reference frame. This track-specific referencing produces inconsistencies and misalignments between orbits that have largely been overlooked *(Lazecký et al., 2023)*. When burst overlap regions cluster predominantly in the far field of faults, ESD corrections can artificially remove true displacement gradients, yielding biased, zero-mean displacement fields. Such biases compromise the ability to monitor plate-boundary deformation over large areas, where preserving both absolute displacement and high spatial resolution across multiple orbits is essential.

Previous BOI studies have demonstrated the value of Sentinel-1 TOPS burst-overlap observations by recovering along-track deformation that is poorly resolved by conventional InSAR. *Grandin et al.* (2016) used burst-overlap and across-track Sentinel-1 interferometry to resolve the 3-D coseismic displacement field of the 2015 $M_w$ 8.3 Illapel earthquake with sub-decimetric accuracy. *Li et al.* (2021) showed that BOI time-series analysis can recover millimeter-per-year interseismic motion across the southern Dead Sea Fault. *Li et al.* (2024) further applied BOI time series to the northern Dead Sea Fault to address slip-rate discrepancies, and *Ma et al.* (2022) tested an improved BOI time-series workflow over the Chaman Fault, integrating phase-SNR enhancement, strain-model filtering, and misregistration correction to recover interseismic deformation and provide constraints on the fault slip rate. More recently, *Nergizci et al.* (2024) applied Sentinel-1 azimuth offset tracking and BOI to the 2023 $M_w$ 7.8 and $M_w$ 7.6 Kahramanmaraş earthquake doublet, using azimuth offset tracking to guide BOI unwrapping and recover meter-scale along-track coseismic displacements across the rupture zone. These studies provide an important foundation for BOI-based earthquake-cycle deformation analysis, but remaining challenges include multi-track consistency, residual orbit/misregistration biases, reference-frame stability, fading-signal contamination, and systematic uncertainty validation.

Here we present a Multi-track Time-Series Burst-Overlap Interferometry (MTSB) method that extends conventional BOI processing and avoids signal suppression associated with ESD-based coregistration by preserving large-scale deformation signals, correcting misregistration through time-series inversion, estimating orbital errors via periodogram velocity analysis, and unifying all tracks in the ITRF2014 reference frame.

## 2 Methodology

The MTSB method overcomes the limitations of BOI by enhancing phase precision, correcting misregistration, and recovering absolute along-track deformation from Sentinel-1 TOPS data (Fig. 1). It entails: (1) burst-overlap interferogram extraction with Sequential Estimator combined with an eigen-decomposition maximum-likelihood (EMI) phase linking *(Ansari et al., 2017*, *2018)*; (2) time-series inversion and periodogram analysis to model deformation and remove misregistration biases; and (3) closure-phase and strain-guided filtering followed by block-wise periodogram inversion to estimate deformation rates tied to plate motion and fault slip. The Sentinel-1 preprocessing and burst-overlap interferogram generation were performed using the CtSent workflow. The main CtSent parameter settings used in this study are summarized in Table S2 to facilitate reproducibility.

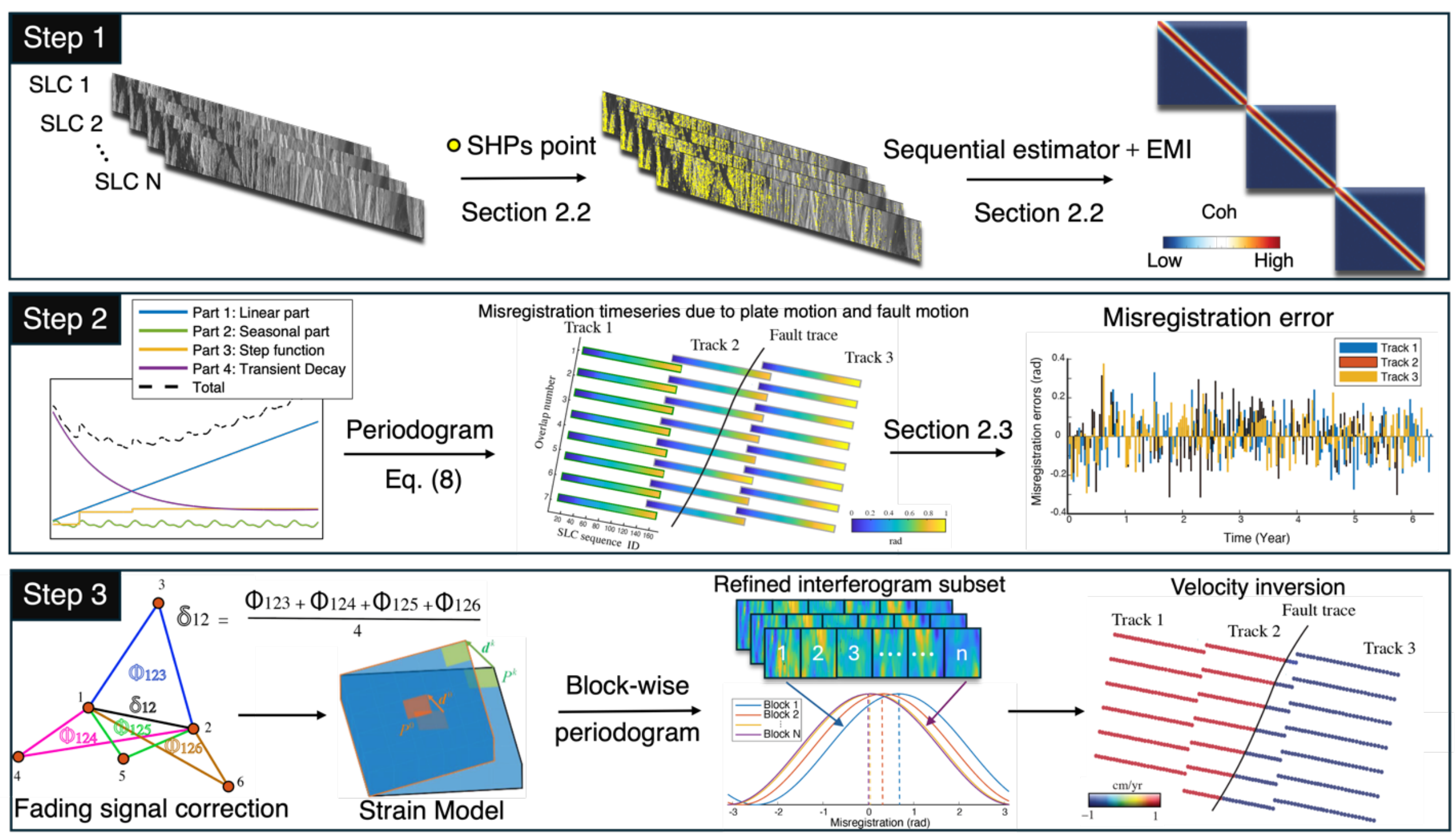

**Figure 1.** Workflow of the proposed methodology. Step one applies the Sequential Estimator and EMI (Section 2.2); step two mitigates misregistration biases through time-series inversion (Section 2.3); and step three estimates deformation rates or displacements after correcting for closure-phase errors (Sections 2.4 and 2.5).

### 2.1 BOI

BOI measures along-track displacements by double-differencing interferometric phases across consecutive overlaps. The phase bias in the presence of azimuth misregistration is expressed as:

$$\Phi_{BOI} = 2\pi \, \Delta f_{ovl} \, \Delta t \tag{1}$$

where $\Delta t$ is the azimuth time error, $\Delta f_{ovl}$ is the spectral separation of forward- and backward-looking images, with each phase cycle corresponding to a displacement range of -65 cm to 65 cm. For each burst overlap region, we can estimate this phase bias by directly differentiating two burst overlap interferograms from two radar echoes *(Prats-Iraola et al., 2012; Yagüe-Martínez et al., 2016)*:

$$\Phi_{BOI} = (r_1 \cdot s_1 *) \cdot (r_2 \cdot s_2 *) * \quad (2)$$

where $r$ and $s$ respectively represent the reference and secondary burst overlap image, $*$ means the complex conjugation operation. Further, (2) can be extended into four phase components: orbital error $\Phi_{orbit}$, along-track deformation phase $\Phi_{def}$ including plate motion and fault motion, ionospheric phase delay $\Phi_{iono}$, and phase noise $\Phi_{noi}$ are expressed as:

$$\Phi_{BOI} = \Phi_{orbit} + \Phi_{def} + \Phi_{iono} + \Phi_{noi} \quad (3)$$

The 3-D positioning uncertainty of Sentinel-1 precise orbits is about 5 cm at the 1-sigma level *(Sansosti et al., 2006)*, which corresponds to ~0.003 pixels of azimuth misregistration error assuming comparable 3-D uncertainties and a 14 m azimuth pixel spacing. Achieving the sub-pixel coregistration accuracy demanded by Sentinel-1, however, requires limiting the azimuthal misregistration error to ~0.001 pixel *(Prats-Iraola et al., 2015; Yagüe-Martínez et al., 2016)*. Ionospheric delay can introduce a nonlinear azimuth-phase bias, though it is often absent and typically suppressed by temporal low-pass filtering. In contrast, decorrelation noise severely impacts BOI phase precision: for a 20 × 4 multi-looked interferogram, the along-track displacement uncertainty is ~1.6 cm at a high coherence of 0.8 but worsens to ~5 cm when coherence decreases to 0.4 (See Text S1).

### 2.2 Stable-scatterer selection and phase accuracy enhancement

Burst-overlap interferograms experience decorrelation from the differing look angles of forward- and backward-looking bursts, lowering phase coherence. To enhance the phase signal-to-noise ratio (SNR), we first identify spatially stable, highly coherent pixels using the Amplitude Dispersion Index (ADI), which is the ratio of amplitude standard deviation to its mean over the time series *(Ferretti et al., 2011; Hooper et al., 2007)*. The ADI is computed from the combined stack of forward- and backward-looking bursts, and pixels satisfying the prescribed threshold are retained as common stable-scatterer candidates. For each retained stable scatterer, we then identify statistically homogeneous neighboring pixels (SHPs) within a local search window using statistical tests of the amplitude time series. These selected SHPs form a local set of statistically similar samples used for covariance estimation in the Seq + EMI phase-linking step.

This two-stage procedure separates stable-scatterer selection from SHP selection and provides sufficient statistically similar samples for robust covariance estimation while accounting for the angular offset between the forward- and backward-looking burst geometries.

Next, we employ a multi-strategy approach integrating Sequential Estimator (Seq) + EMI (Eigen-decomposition-based Maximum Likelihood Estimator of Interferometric phase) *(Ansari et al., 2017, 2018; Ma et al., 2022)*. The EMI enhances computational efficiency and approaches the Cramér-Rao lower bound on phase accuracy by estimating the interferometric phase from the principal (minimum-eigenvalue) eigenvector of the full covariance matrix *(Ansari et al., 2018)*. Its formulation is defined as:

$$\hat{\Sigma} = \frac{1}{N}\sum PP^{H} \tag{4}$$

$$\hat{\Phi} = argmin_{\Phi}\left\{\theta^{H}\left(\Upsilon^{-1} \circ \hat{\Sigma}\right)\theta\right\} \tag{5}$$

where $P$ denotes the set of pixels within the corresponding SHP neighborhood. The superscript $H$ denotes the Hermitian, or conjugate, transpose. $\hat{\Sigma}$ is the covariance matrix estimated from the pixels in the corresponding SHP neighborhood, $\hat{\Phi}$ is the estimated phase series, and $\theta$ is the minimum eigenvector of the Hadamard product of $\Upsilon^{-1} \circ \hat{\Sigma}$. $\Upsilon = \left|\hat{\Sigma}\right|$ represents the real coherence matrix in which $\gamma_{ij}^{2k}$ is the coherence between two acquisitions estimated from forward and backward-looking subsets:

$$\Upsilon = \begin{bmatrix} 1 & \cdots & \gamma_{1N}^{2k} \\ \vdots & \ddots & \vdots \\ \gamma_{N1}^{2k} & \cdots & 1 \end{bmatrix} \tag{6}$$

$$\gamma_{ij}^{2k} = \sum_{P\in\Omega} exp(\sqrt{-1}\cdot\left[arg(P_{ij}^{up}P_{ij}^{up*} - \Phi_{local}^{up} + P_{ij}^{lo}P_{ij}^{lo*} - \Phi_{local}^{lo})\right]) \tag{7}$$

$\Phi_{local}^{up}$ and $\Phi_{local}^{lo}$ respectively represent the phase associated with flat-earth and topography in the forward- and backward-looking interferograms. EMI enhances phase SNR by incorporating multiple samples from both subsets, reducing singularity of covariance matrices *(Ma et al., 2022)*. Seq further reduces the risk of matrix singularity by partitioning the full covariance matrix into block diagonals and sequentially compressing each into a rank-1 subspace *(Ma et al., 2020)*.

Through linking phase information of these subspaces, these block diagonals are tied to a common phase datum, enabling the estimated phase to attain a higher SNR than with the original EMI. In this study, the block-diagonal size in the Sequential Estimator (Seq) was set to 10, meaning that each covariance submatrix contains 10 acquisitions during sequential phase estimation rather than representing a spatial block dimension. This empirical value follows *Ma et al.* (2022), where it was successfully applied.

### 2.3 Misregistration error estimation and correction

BOI phase errors arise from multiple sources: orbital errors, which behave systematically over time, and random contributions such as ionospheric or geometric effects. Orbital errors are particularly problematic, as they shift displacement estimates into an unknown reference frame and result in inconsistencies when combining data from multiple orbits. For Sentinel-1, with a maximum spectral separation of ~5 kHz, even a 1/100-cycle phase ramp (3.6°) produces an azimuth misregistration of ~0.0009 pixel (~1.3 cm), comparable to surface displacements from plate motion.

To estimate and correct the misregistration errors, we first apply a periodogram-based spectral analysis to each burst-overlap interferogram, identifying the dominant phase shift due to misregistration or deformation *(Yagüe-Martínez et al., 2016)*, as shown below

$$\widehat{\Phi}_{mis} = argmax_{\Phi_{mis}} \left[ \Re \left( \sum_{p \in \Omega} e^{\sqrt{-1}(\Phi_{BOI} - \Phi_{mis})} \right) \right] \tag{8}$$

where $\Omega$ denotes each burst overlap, argmax [·] denotes the argument of the maximum, and $\Re$ denotes the real part of a complex number. $\Phi_{BOI}$ represents the phases in the burst-overlap interferogram. $\Phi_{mis}$ is the modeled phase for a single burst-overlap interferogram. It represents the dominant phase shift in the burst-overlap interferogram caused by misregistration, which may stem from errors or along-track deformation in Eq. 3. The accuracy of the displacement is correlated with coherence and the number of involved points; thus, precision can be improved with increasing coherence and the number of averaged observations. Thousands of observations within each overlap enable the accurate detection of a constant phase value that maximizes the periodogram's real-valued sum in Eq. 8. Next, we invert for the misregistration time-series from the interferogram subset, then we model time-dependent ground deformation using a comprehensive parameterization that captures geophysical processes, including plate motion, fault motion, fault related (co-, inter-, and postseismic) deformation, and non-tectonic effects. The model we use is constructed by adding linear velocity, seasonal variation, a Heaviside function centered at $T_i$, and a B-Spline part into velocity inversion; the mathematical formulation of this model is expressed as:

$$
\begin{aligned}
D_p(t) &= V_p t + \tau \\
&+ \sum_{i=1}^{2} a_i \sin(\omega_i t) + b_i \cos(\omega_i t) \\
&+ \sum_{i \in \Omega} \Delta i(\Omega) \mathcal{H}(t - T_i) \\
&+ \sum_{i \in \Omega} c_i B^n\left(t - t_i^{ctrl}\right) + d_i B_{\int}^n\left(t - t_i^{ctrl}\right)
\end{aligned}
\tag{9}
$$

The linear term $V_p t$ represents steady deformation, such as long-term plate motion or interseismic fault slip, where $V_p$ is the secular deformation rate, and $\tau$ is the constant reference offset of the displacement time series. The seasonal component contains annual and semiannual sinusoidal terms, with $\omega_1 = 2\pi\ yr^{-1}$ and $\omega_2 = 4\pi\ yr^{-1}$ ; $a_i$ and $b_i$are the corresponding sine and cosine coefficients. The step component represents prescribed abrupt offsets, such as known coseismic displacements or equipment changes, where $\Delta i(\Omega)$ is the offset magnitude and $T_i$ is the event epoch. The B-spline basis is adopted as a flexible empirical representation of transient deformation, allowing different temporal evolution patterns to be approximated without assuming a specific physical relaxation law. In this study, cubic B-spline basis functions with a 3-month knot spacing are used only for the Türkiye postseismic case. Temporal components are activated only when supported by the expected deformation behavior, and the complete case-specific parameterizations are summarized in Table S3. Sensitivity tests show that introducing additional basis functions does not alter the recovered secular velocity (Fig. S1).

To implement this case-dependent parameterization, the deformation and residual misregistration contributions are separated through the following time-series procedure. First, the burst-overlap interferogram network is inverted to obtain an epoch-wise phase time series for each overlap region. Each time series contains both deformation-related phase and residual misregistration. The case-specific temporal model in Equation (9) is then fitted to each time series using least squares, where the linear, seasonal, step, and transient components (as applicable) represent the modeled deformation and other prescribed temporal signals. The remaining residuals are assumed to be dominated by residual misregistration. For each epoch, the residuals are spatially averaged across all burst-overlap regions within the same track to estimate an epoch-wise, track-specific misregistration term, which is subsequently removed from the original burst-overlap interferograms.

### 2.4 Fading signal correction

The fading signal, a systematic velocity bias caused by multi-looking of interferograms, is well documented in conventional InSAR processing *(Ansari et al., 2021; Ma et al., 2025; Maghsoudi et al., 2022)*, but its impact on BOI has not previously been examined. Here we confirm its presence in BOI and propose a correction to enhance phase robustness (see Discussion). This fading signal originates from phase inconsistencies among interferometric triplets ($\Phi_{ijk} \neq 0$ in Eq. 10), which degrade velocity estimates. To mitigate this effect, we apply closure phase analysis: by enforcing phase consistency across interferometric triplets and stacking a network of closure phases, the fading signal for each interferogram can be isolated and removed *(Ma et al., 2025)*, as expressed in Eq. 10.

$$\Phi_{ijk} = \Phi_{ij} + \Phi_{jk} - \Phi_{ik} \tag{10}$$

$$\delta_{ij} = \frac{1}{m}\sum_{k=1}^{m} \Phi_{ijk} \tag{11}$$

$\Phi_{ijk}$ denotes the closure phase of the interferometric triplet formed by $\Phi_{ij}$, $\Phi_{jk}$, $\Phi_{ik}$ where $i$, $j$, $k$ correspond to the acquisition epochs. For a given interferogram $\Phi_{ij}$, $m$ is the number of valid interferometric triplets containing that interferogram. The systematic fading-phase contribution $\delta_{ij}$ is estimated by stacking the closure phase contributions from these m valid triplets. Because the target interferogram $\Phi_{ij}$ is common to all $m$ triplets, its fading-phase contribution is repeatedly represented in the closure phases, whereas contributions from the other interferograms vary among triplets and are reduced through stacking. The average closure phase therefore provides an estimate of the fading-phase bias $\delta_{ij}$ associated with $\Phi_{ij}$, which is subsequently subtracted from the corresponding BOI interferogram. By removing the fading signal, our method reduces systematic fading-related phase biases in the reconstructed interferograms.

### 2.5 Strain model

Following fading-signal correction, we apply a strain-based regularization model (SM) that exploits the spatial autocorrelation of interferometric phase to further boost SNR. Because SHP-based phase linking alone may discard valid observations, this method "recovers" additional pixels by enforcing the kinematic consistency of neighboring phase measurements. The approach assumes that the surface displacement field is locally continuous and can be approximated by a first-order (linear) strain tensor. Thus, the 3-D displacement vector at any point is expressed as a linear function of its position relative to surrounding reference points. For BOI interferograms where deformation is predominantly along the azimuth, we reduce the full 3-D strain formulation to a 1-D representation. In this simplified case, the along-track displacement $d^k$ at location $P^k$, with respect to a reference point, $P^0$ satisfies:

$$d^k = d^0 + H \cdot \Delta x^k \quad (12)$$

where $d^0$ is the displacement vector at the reference point $P^0$, $d^k$ is the displacement at point $P^k$, $\Delta x^k = x^k - x^0$ is the spatial offset, and $H = \partial d / \partial x$ is the displacement gradient. This linear formulation assumes a locally homogeneous strain field in which adjacent points deform coherently according to their spatial offsets. As illustrated in Fig. 1, the method enforces this continuity by fitting a local linear strain field that links the displacement estimates of neighboring pixels. A linear observation matrix is assembled from the relative positions and measured phases of multiple adjacent points, and the reference displacement and strain parameters are then estimated by weighted least squares. This procedure suppresses high-frequency phase noise while preserving long-wavelength tectonic signals, thereby improving the stability of the central-point phase estimate. The noise-filtered interferograms are subsequently partitioned into n × n km spatial tiles, and the periodogram in Equation (8) is applied independently to each tile. This block-wise implementation produces dense local phase estimates and therefore improves the spatial resolution of the retrieved along-track deformation field.

To improve reproducibility, we summarize the main MTSB-specific processing parameters, including stable-scatterer selection, Seq + EMI phase linking, misregistration correction, fading-signal correction, strain model, and block-wise periodogram settings, in Table S3. Common settings were retained wherever the same processing module was applied, whereas case-specific choices (e.g., temporal parameterization) are explicitly identified.

## 3 Synthetic data tests

To assess the performance of MTSB, we simulate 40 single-look complex (SLC) acquisitions with a 12-day revisit interval. Each interferogram phase is modeled as the sum of a linear deformation signal, long-wavelength background contributions, and decorrelation noise. The synthetic velocity field, generated from a 1-D dislocation model with a maximum rate of 1 cm/yr, includes two zones of sharp displacement gradients (Fig. 2), representative of common natural settings. Each burst-overlap SLC image has dimensions 1456×12311 pixels (azimuth × range) and is multi-looked by 4×20 to form the BOI interferograms. Noise phases are drawn from the prescribed coherence matrix. Topographic and flat-Earth phase components are simulated using precise orbit ephemerides and an external Digital Elevation Model (DEM).

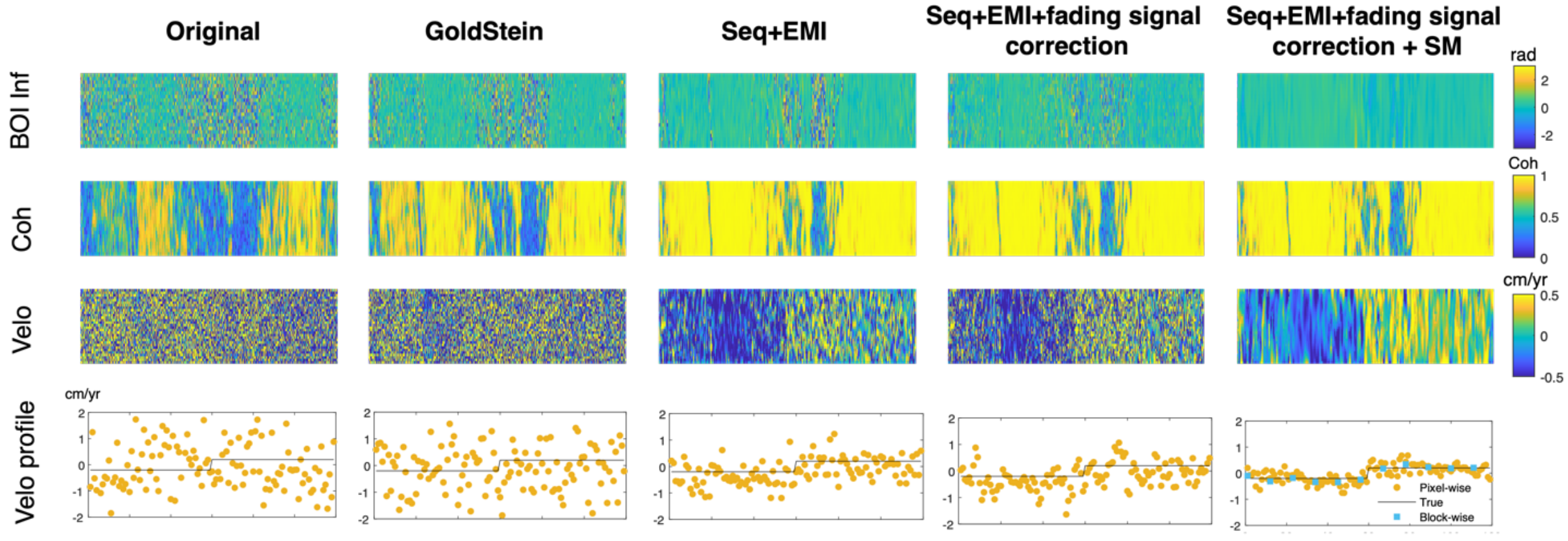

**Figure 2.** Burst-overlap interferograms after phase estimation and the related inverted velocity. The first row shows the original, Goldstein-filtered, Seq + EMI, Seq + EMI + fading-signal correction, and Seq + EMI + fading-signal correction + strain model phases, illustrating a progressive increase in SNR from left to right. The second row shows the calculated coherence of the first row. The third row shows the calculated velocity for each method. The fourth row

presents the comparison between the velocity profiles and the simulated velocity, with pixel-wise and block-wise comparisons shown in the last velocity profile.

We next implement the workflow shown in Fig. 1 to retrieve all burst-overlap interferograms. The unfiltered wrapped phase and the Goldstein-filtered *(Goldstein and Werner, 1998)* phase (16-pixel square window, $\alpha = 1$, eight zero-padding iterations) are used as benchmarks. Our approach reconstructs the BOI phase by first correcting the fading signal and then applying the strain-based model to the Seq-enhanced EMI output. The first and second rows of Fig. 2 present the 60-day wrapped interferograms and their corresponding coherence, where the SNR of the BOI phase increases progressively from left to right.

Fig. 3 evaluates the accuracy of the velocity field as a function of the number of images and the number of neighboring pixels used in the block-wise strain estimation. For the analysis, we extended the time-series length to 200 images, with an average coherence of 0.4, considering only decorrelation noise as expressed in Text S1. Enlarging the neighborhood improves velocity precision, but overly large blocks reduce the density of independent observations. Balancing these effects, we adopt a 2.5 km block size and 120 nearest neighbors for the strain model in the real-data tests.

This hybrid strategy combines dense spatial sampling with strong statistical robustness: by stacking hundreds of observations within each block, it yields a marked accuracy gain relative to conventional pixel-wise BOI methods. The improvement is quantitatively and visually confirmed by the simulated velocity profiles in Fig. 2.

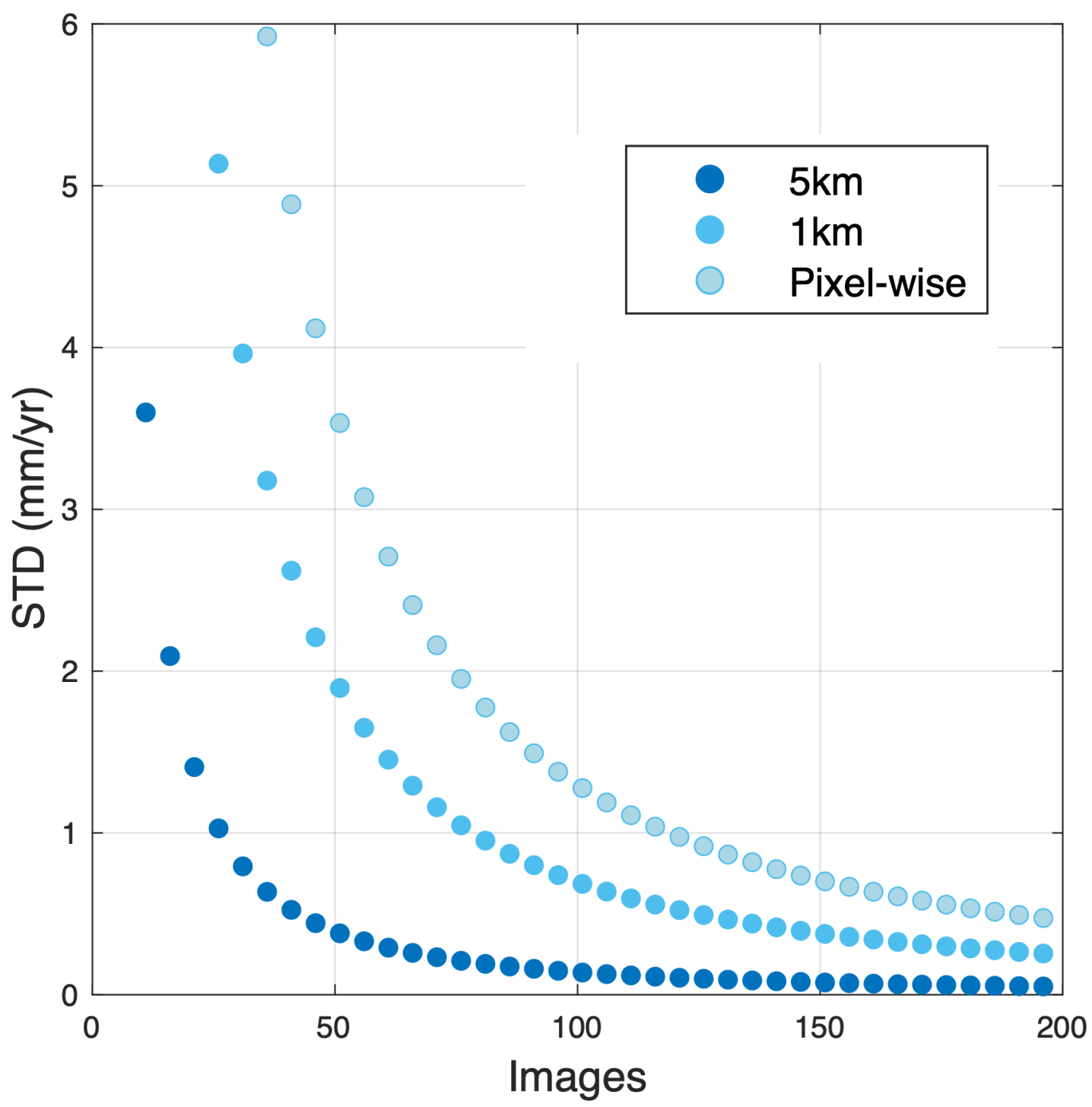

**Figure 3.** Theoretical accuracy improvement with block-wise method (block sizes 1 km and 5 km) compared with pixel-wise method.

## 4 Applications of MTSB Across the Earthquake Cycle

We evaluated MTSB using three case studies spanning interseismic deformation along the southern Dead Sea Fault, postseismic deformation following the 2023 Türkiye earthquake doublet, and coseismic displacement associated with the 2015 $M_w$ 8.3 Illapel earthquake. Together, these applications test the method across signals with different amplitudes, temporal characteristics, and spatial scales.

### 4.1 Interseismic deformation along the Dead Sea Fault in the Gulf of Aqaba

The Dead Sea Transform Fault, a major sinistral plate boundary separating the Sinai microplate and Arabian plates, extends approximately 1,000 km from the Maras Triple Junction in southern Türkiye to the Red Sea Rift (Fig. 4). In its southernmost segment, the Gulf of Aqaba, geological records dating back to the Early Miocene indicate approximately 105 km of cumulative left-lateral offset *(Freund et al., 1970; Garfunkel et al., 1981; Quennell, 1958)*. Contemporary geodetic studies estimate a slip rate of 4.3-5.0 mm/yr, supported by geological markers *(Sadeh* *et al., 2012)*, paleoseismic trenches *(Hamiel and Piatibratova, 2021)*, Global Navigation Satellite System (GNSS) networks *(Castro-Perdomo et al., 2022; Gomez et al., 2020)*, and BOI results *(Li et al., 2024)*. However, large-scale horizontal displacement mapping remains challenging, as conventional BOI struggles to resolve spatially continuous deformation fields due to its wide deformation zone (>100 km) and limited GNSS stations. To address this, we applied the proposed MTSB method to retrieve the horizontal displacement field using multi-track data, processing Sentinel-1 SAR images acquired from 2014 to 2021 along descending tracks 21, 94, and 123 (Fig. 4 and Table S4). This observation period was selected to match that used by Li et al. (2021), thereby enabling a direct comparison between the conventional BOI and MTSB results. We performed geometric coregistration of all images using the Coregistration Toolbox for Sentinel-1 (CtSent) software with the Shuttle Radar Topography Mission (SRTM) 30-m DEM and precise orbit data. Following coregistration, we extracted 36 burst overlaps from the three tracks and constructed a small baseline BOI interferogram network for each overlap.

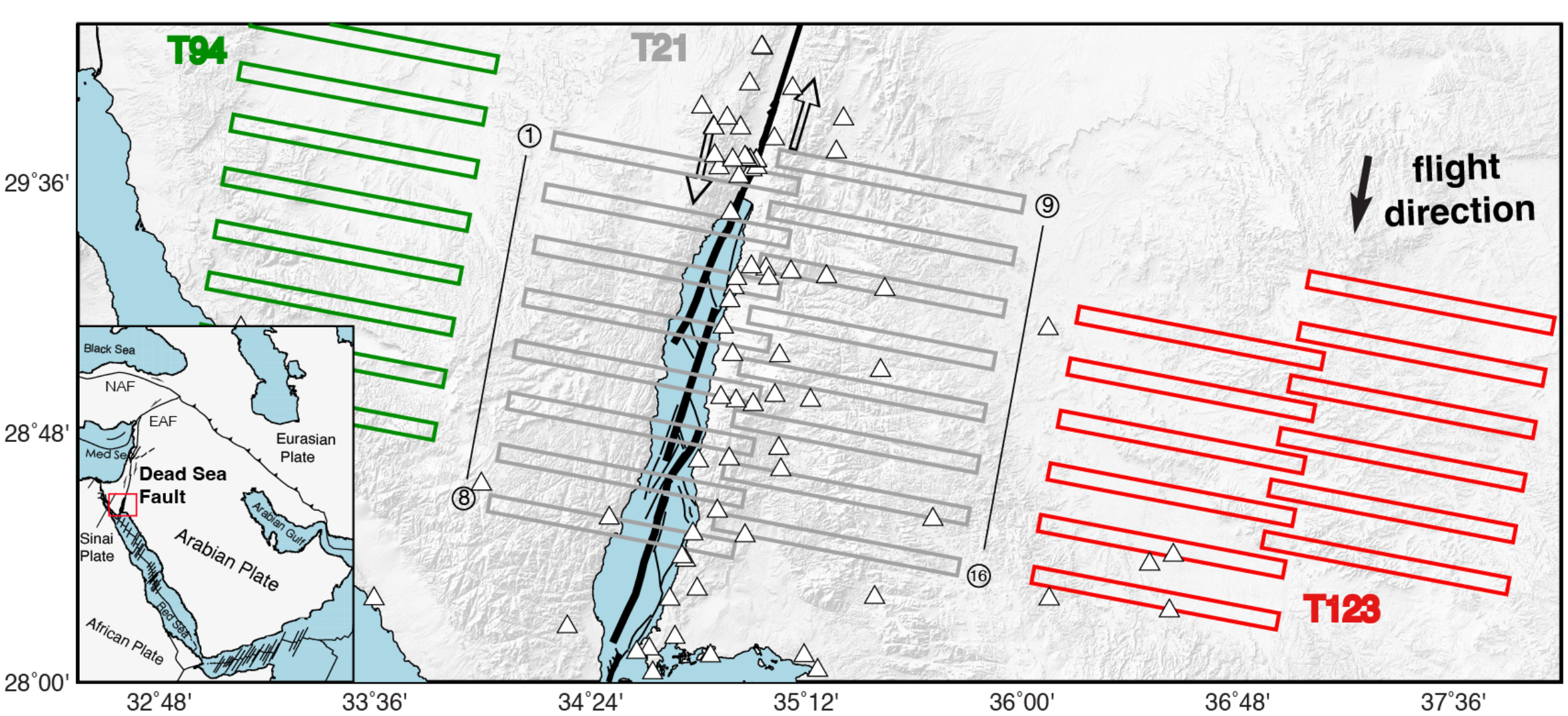

**Figure 4.** Map of the study region, showing the southern Dead Sea Fault trace as a solid black line, and green, gray, and red rectangles indicating the burst-overlap regions from tracks T94, T21, and T123, respectively.

To enhance phase accuracy and correct errors in the BOI interferograms, we applied the phase optimization and error correction procedures described in Section 2. Fig. 5 illustrates representative burst-overlap interferograms that illustrate the effectiveness of the proposed workflow. The first, second, and fifth columns of Fig. 5a show the unwrapped original, Goldstein-filtered, and MTSB-refined BOI interferograms with temporal baselines of 60 and 120 days. As demonstrated in the synthetic tests, our method significantly reduces decorrelation noise in the BOI phase, resulting in clearer patterns. The bias induced by the fading signal is evident in individual interferograms and cannot be neglected in BOI processing. After applying the strain-based spatial filter, the interferograms exhibit reduced noise, and additional coherent points are successfully recovered.

Before fading-signal correction, azimuth misregistration is estimated for each burst-overlap region using a periodogram analysis across the three tracks. For each overlap, the BOI interferogram subset provides the initial misregistration estimate, which is then used to invert a misregistration time series by least-squares optimization (Fig. 5). Time series for all three tracks are jointly inverted (Track T21 in Fig. 5b). The case-specific temporal parameterizations are summarized in Table S3. For the Dead Sea interseismic analysis, the seasonal terms are treated as nuisance components representing the observed periodic variability, rather than as seasonal tectonic deformation. After removing the fitted deformation components, including the linear trend and seasonal nuisance terms, the residual misregistration is obtained (Fig. 5c). For example, the residuals from overlaps Ovl1–Ovl16 in track T21 exhibit epoch-dependent common phase offsets that are largely attributable to orbital errors *(Ma et al., 2022)*. After removal of the modeled deformation, these residuals fluctuate around a nearly constant, near-zero mean level without a systematic temporal trend (Fig. 5c). These residuals are subsequently corrected within the interferogram subsets to ensure a consistent displacement field across tracks. The separation relies on the different characteristics of the two terms: deformation is represented by the case-specific time-series model, whereas residual misregistration is estimated from the remaining epoch-wise residuals through spatial averaging across all burst-overlap regions within each track. We evaluated potential leakage by repeating the time-series inversion with and without the residual misregistration correction and comparing the resulting deformation estimates (Table S5). The correction produces only a small change in the estimated secular velocity, with a mean corrected-minus-uncorrected difference of −0.036 cm/yr and a maximum absolute difference of 0.084 cm/yr. The small nonzero mean indicates that the correction has only a limited influence on the recovered secular deformation, although a minor systematic effect cannot be excluded. In addition, the residual misregistration time series in Fig. 5c fluctuate around zero without a clear long-term trend. Together, these results suggest that the correction primarily removes epoch-dependent common-mode phase biases and has only a limited influence on the recovered secular velocity, rather than demonstrating the complete absence of deformation-signal leakage.

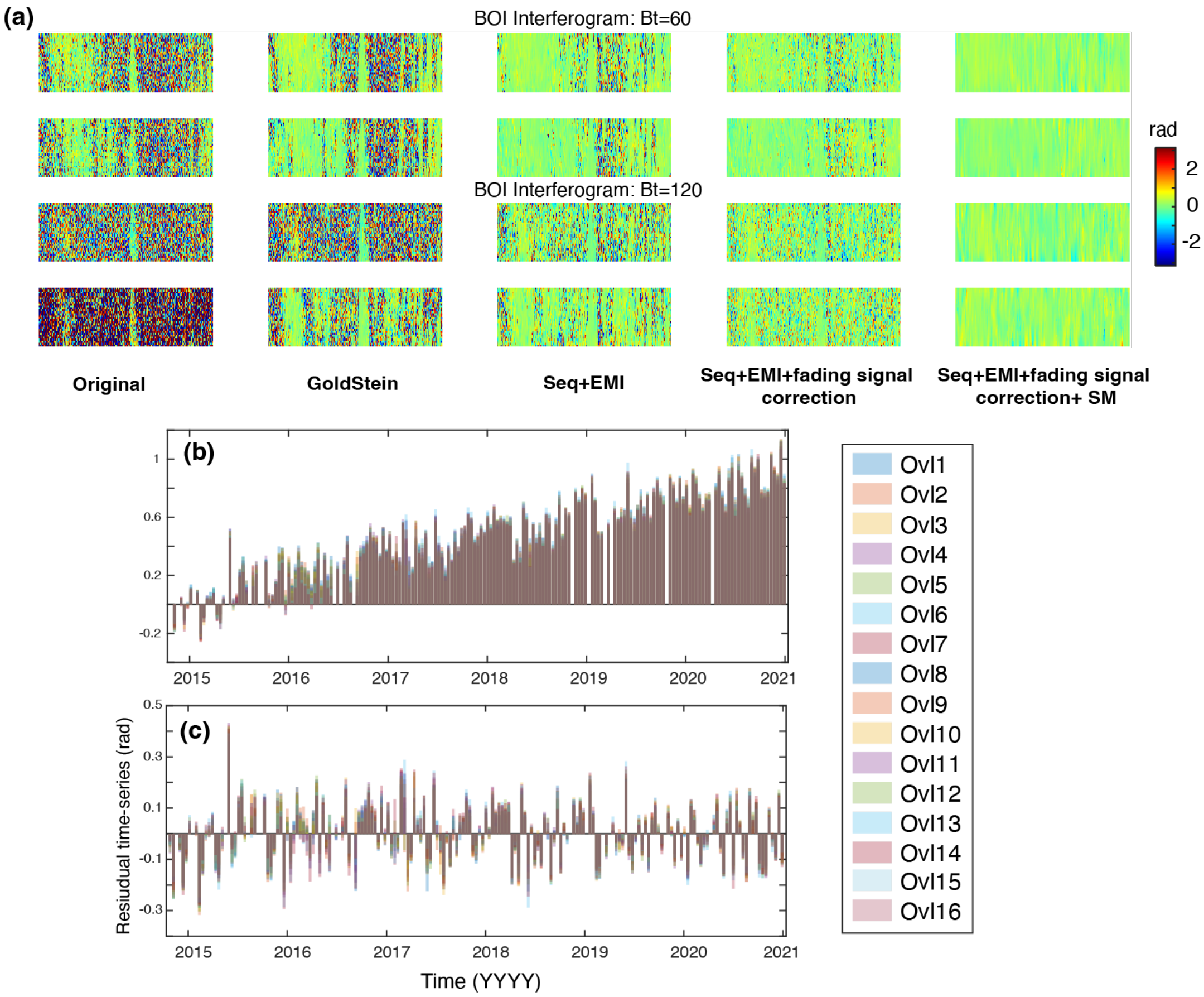

**Figure 5.** (a) Burst-overlap interferograms after phase-accuracy enhancement. The first and second columns show the original and Goldstein-filtered interferograms, respectively, whereas the third to fifth columns illustrate the sequential improvements from the proposed MTSB workflow. (b) Inverted epoch-wise misregistration time series for all overlap regions (Ovl1-Ovl16) in Sentinel-1 track T21. The horizontal axis in (b) and (c) represents the acquisition time of each Sentinel-1 image; the epoch-wise misregistration estimates are obtained by time-series inversion of the burst-overlap interferogram network. (c) Residual misregistration time series after removing the fitted deformation/misregistration model from (b). Colors denote different burst-overlap regions, Ovl1-Ovl16, whose locations are shown in Fig. 4.

After misregistration correction, the fading signal is estimated following Eq. 11 and removed. Finally, the strain model is applied to enforce spatial continuity and further enhance the SNR. A statistical comparison between our reconstructed interferogram subsets and those filtered with the conventional Goldstein filter across all overlap regions is presented in Fig. 6, where SNR is defined as:

$$SNR = 10\log_{10}\frac{\sigma_{\Phi_{boi}}}{\sigma_{\hat{\Phi}_{boi}}} \tag{13}$$

where $\sigma_{\Phi_{boi}}$ and $\sigma_{\hat{\Phi}_{boi}}$ represent the phase standard deviation of the original and refined BOI interferograms *(Ma et al., 2022)*. A higher SNR value indicates improved phase quality. The radial bar chart in Fig. 6 highlights a substantial increase in SNR following the application of our proposed MTSB method. Compared to the Goldstein-filtered results, the MTSB algorithm significantly reduces the phase standard deviation across all overlap regions. For the Dead Sea Fault interseismic case, the phase standard deviation using MTSB typically ranges from 0 to 0.4 rad, while that of the Goldstein method spans from 0.1 to 1 rad. This considerable reduction in phase noise confirms that MTSB enhances phase accuracy, thereby enabling more reliable detection of subtle interseismic deformation signals along the fault zone.

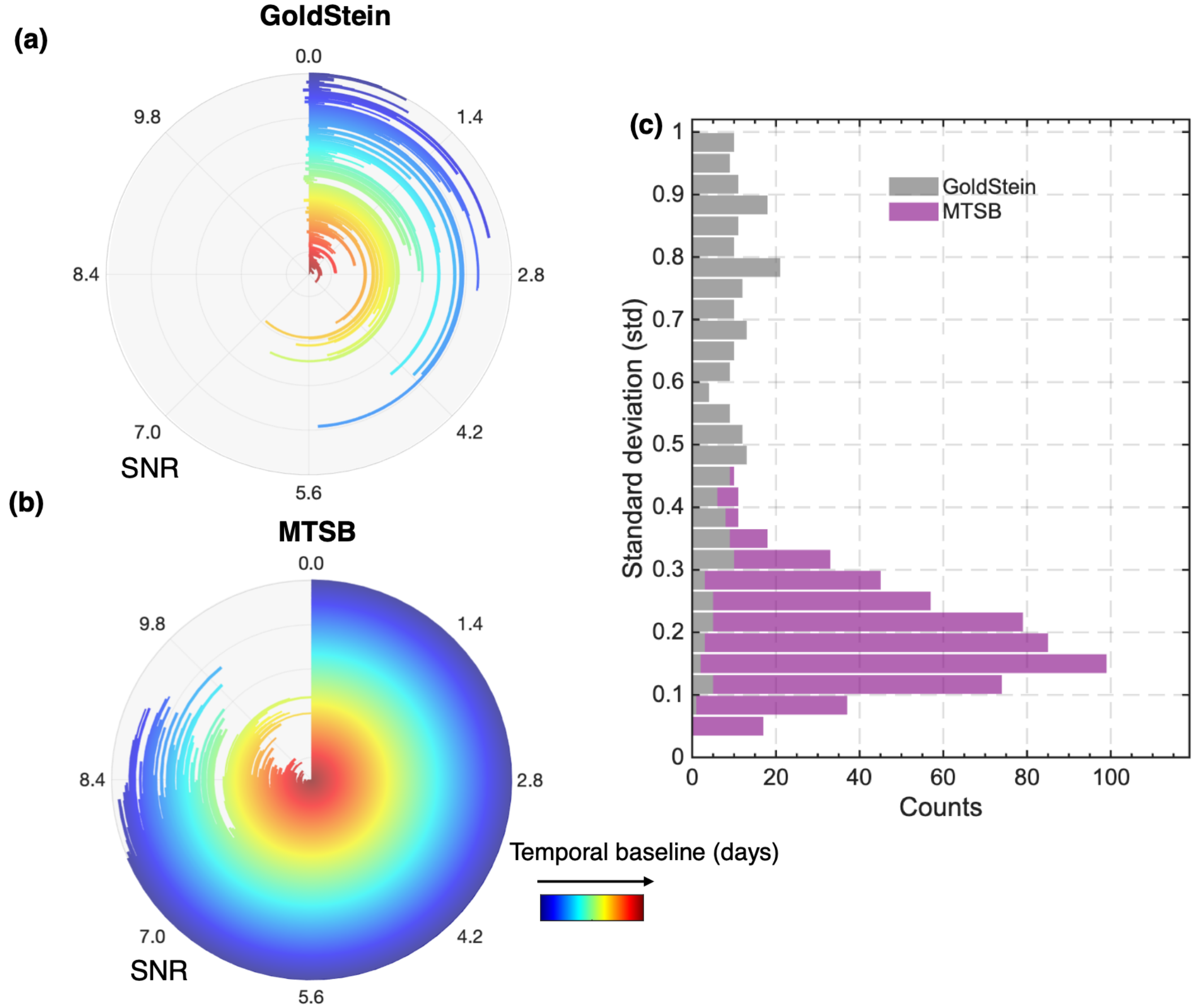

**Figure 6.** Quantitative comparison of phase quality for the Dead Sea burst-overlap interferograms processed with conventional Goldstein filtering and the proposed MTSB workflow. (a, b) Polar plots of phase signal-to-noise ratio (SNR) for the Goldstein-filtered and MTSB-refined interferogram subsets, respectively. The SNR statistics in (a) and (b) were computed from five burst-overlap regions, and colors indicate temporal baseline in days. Larger radial values indicate higher phase SNR. (c) Histograms of phase standard deviation for the conventional Goldstein-filtered results (gray, Classic) and MTSB-refined results (purple, New), computed over valid coherent pixels from all overlap regions. The shift toward lower phase standard deviation and higher SNR demonstrates improved phase precision after MTSB processing.

After correcting the BOI interferogram subset, we applied a block-wise periodogram function using a 2.5 km × 2.5 km window, which yielded a spatially continuous interseismic displacement field with sufficient observational density. A comparison between the conventional method and the proposed MTSB approach (Fig. 7) highlights the substantial improvements achieved by our method over traditional BOI processing. The derived velocity field attains a spatial resolution of 2.6 km, over an order of magnitude finer than that of regional GNSS networks, while maintaining sub-centimeter-per-year accuracy through time-series analysis (Fig. 7). To quantify the formal $1\sigma$ velocity uncertainty of the MTSB-derived interseismic velocity estimates, we propagated the full covariance matrix of the epoch-wise displacement estimates through a generalized least-squares linear-rate model. The spatial distribution of the resulting formal velocity uncertainty is shown in Fig. S2, with most MTSB observations exhibiting uncertainties of approximately 0.6–1.0 mm/yr. This enhanced resolution allows for the accurate detection of large-scale tectonic motion, and the comparison with GNSS velocities projected into the Sentinel-1 along-track direction yields an RMSE of 2.0 mm/yr and a bias of 1.0 mm/yr, supporting the millimeter-per-year accuracy of the MTSB-derived interseismic velocity field (Fig. S3 and Table S6). Notably, the raw velocity field derived from conventional BOI (Fig. S4) exhibits noticeable discontinuities between adjacent tracks. With our new approach, the track-to-track velocity discrepancy is reduced from 2 mm/yr (using conventional BOI) to 0.7 mm/yr, achieving a uniform displacement field referenced to the ITRF2014 geodetic framework.

To quantify fault kinematics, we projected velocities onto a fault-parallel profile and inverted them using an elastic half-space dislocation model *(Savage and Burford, 1973)*. The fault geometry is aligned with the central axis of the Gulf of Aqaba. The dislocation model yields a slip rate of 4.8 ± 0.3 mm/yr and a locking depth of 8.5 ± 1.2 km (Fig. 7 profile B), consistent with the previous BOI-derived estimates of *Li et al.* (2021) and GNSS-derived estimates *(Castro-Perdomo et al., 2022)*.

In addition to the velocity field, Fig. S5 compares displacement time series derived from conventional BOI and MTSB. The MTSB time series exhibit lower scatter and more coherent temporal evolution, providing a more stable basis for estimating continuous interseismic deformation. These improvements are consistent with the enhanced phase quality and residual-misregistration correction demonstrated in Figs. 5 and 6.

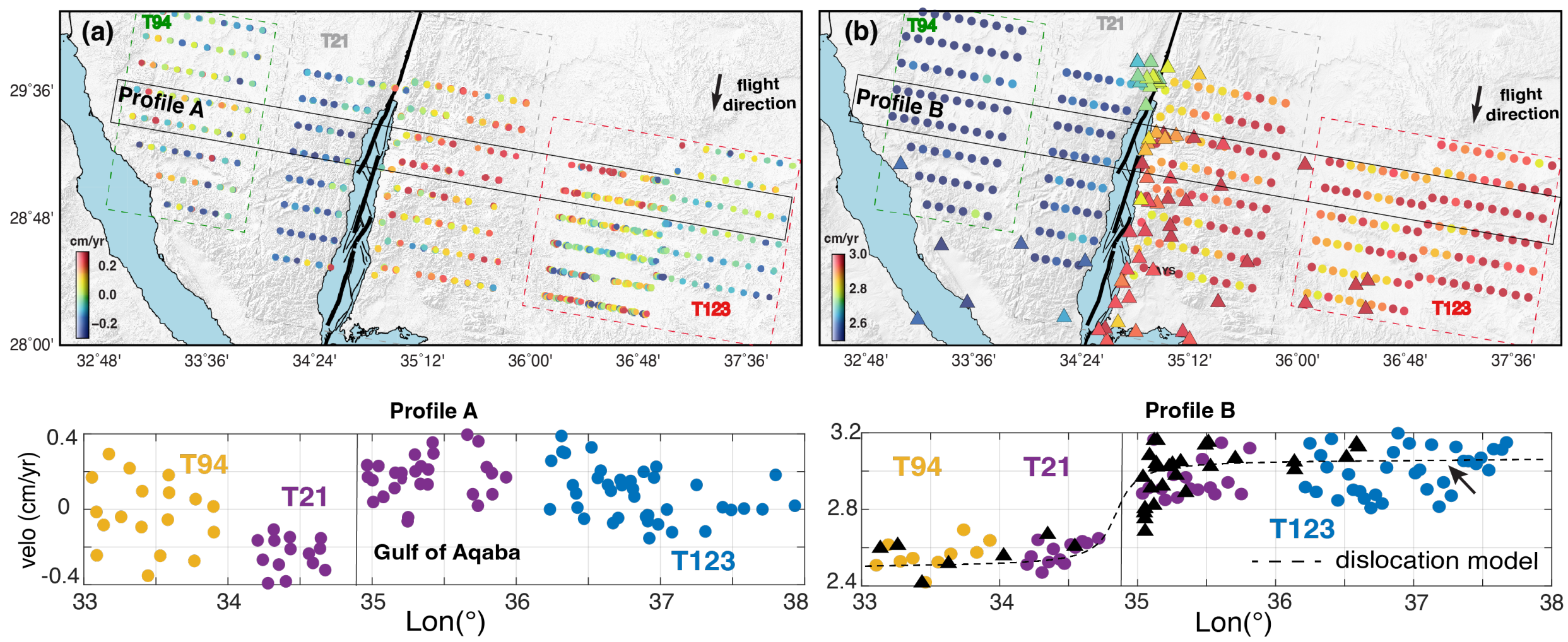

**Figure 7.** Horizontal velocities along the NE12° direction across the southern Dead Sea Fault, derived from (a) the conventional BOI method and (b) the proposed MTSB method. Colored circles show BOI/MTSB-derived velocities, and triangles represent GNSS velocities projected onto the NE12° direction. The observations in (a) are extracted according to the block definitions shown in (b). The lower panels show the corresponding cross-fault velocity profiles along Profiles A and B; black triangles in Profile B indicate the projected GNSS velocities for direct comparison with the MTSB profile. The dashed line in Profile B denotes the dislocation model.

### 4.2 Postseismic deformation of the 2023 Türkiye earthquake doublet

On 6 February 2023, a devastating earthquake doublet ($M_w$ 7.8 and $M_w$ 7.6) struck the tectonically complex region spanning southern Türkiye and northwestern Syria *(Mai et al., 2023)*. The first shock ($M_w$ 7.8) exhibited left-lateral strike-slip motion along a southwest-northeast-trending rupture zone. The sequence ruptured a complex, predominantly left-lateral fault network: the first event propagated along multiple segments of the southwest–northeast-trending East Anatolian Fault system, whereas the second event occurred approximately 9 hr later on the Çardak–Sürgü fault system to the north *(Jia et al., 2023; Reitman et al., 2023)*. To quantify the postseismic deformation, we processed one year of ascending and descending Sentinel-1 SAR datasets covering an approximately 500 km-wide area (Table S7). The processing workflow followed the same procedures described in the interseismic case study.

The displacement fields derived from descending and ascending tracks clearly delineate the horizontal motion across the $M_w$ 7.8 coseismic rupture (black line in Fig. 8a). The postseismic displacements exhibit a pronounced asymmetric pattern along the fault trace, with maximum along-track motion reaching ~8 cm during the first postseismic year. Most of the signal is concentrated south of the main rupture and north of the secondary rupture. Overlapping segments from adjacent tracks along profiles A and B reveal a coherent, spatially continuous displacement field (Fig. 8). This continuity contrasts with the track-dependent reference offsets observed in conventional BOI processing, as illustrated by the Dead Sea comparison in Fig. S4.

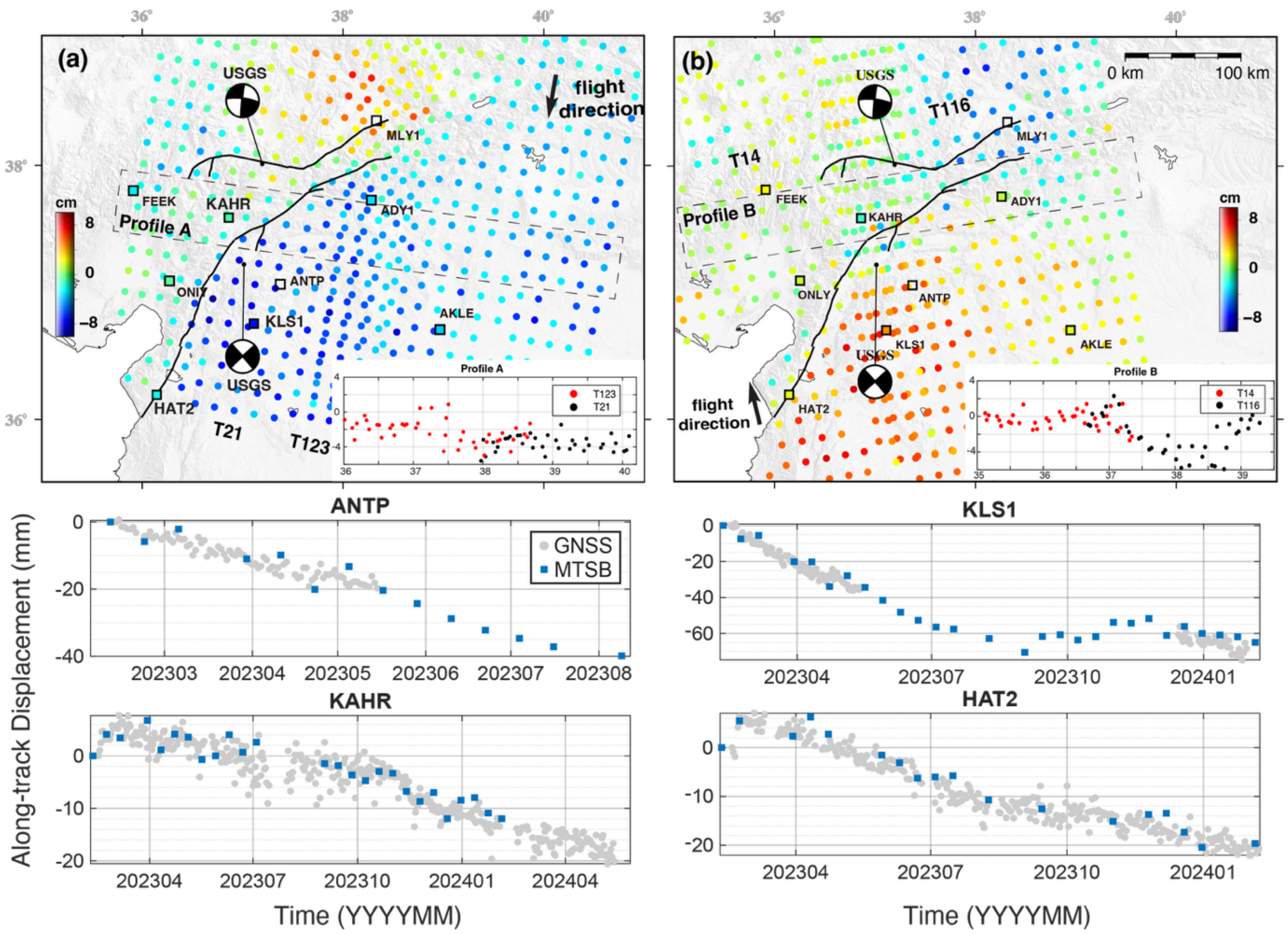

**Figure 8.** Accumulated postseismic along-track displacement fields derived from descending (a) and ascending (b) Sentinel-1 tracks over the approximately one-year, track-specific observation intervals listed in Table S7. Squares indicate GNSS station locations, and their colors represent GNSS displacements projected into the corresponding along-track direction using the same color scale as the MTSB field. Insets show the corresponding cross-profiles for the descending and ascending tracks, respectively. The lower panels present postseismic along-track time series from the descending track, compared with GNSS measurements projected onto the descending along-track direction.

To further validate these results, GNSS time series were projected into the corresponding ascending and descending along-track directions and compared with the MTSB-derived displacements (Fig. 8). For the GNSS–MTSB time-series comparisons, epochs whose residuals relative to the fitted temporal model exceeded three times the median absolute deviation were identified as outliers. The same criterion was applied consistently to all stations and both orbit tracks before calculating the comparison statistics. The GNSS–MTSB comparison yields RMSE values of 0.90 and 1.56 cm and biases of -0.10 and -0.44 cm for the ascending and descending tracks, respectively, demonstrating centimeter-level agreement for the recovered postseismic displacements (Fig. S6 and Table S6). Given the GNSS station coverage, we focus on comparing the time-series evolution rather than absolute velocities (Fig. 8, Fig. S7 and Fig. S8). Both the GNSS and MTSB-derived time series reveal a sustained displacement throughout the approximately one-year observation period.

### 4.3 Coseismic deformation of the 2015 $M_w$ 8.3 Illapel earthquake

Coseismic deformation, characterized by rapid, meter-scale horizontal displacements, plays a critical role in understanding earthquake rupture processes and assessing seismic hazards *(Liu et al., 2022; Simons et al., 2002)*. In contrast to conventional BOI techniques, MTSB provides spatially continuous, ITRF2014-referenced along-track displacement measurements across multiple tracks. While the interseismic case demonstrates the utility of MTSB for resolving subtle deformation, the Illapel earthquake provides a favorable test case for evaluating its performance in a coseismic setting because the offshore rupture produced a relatively smooth displacement field across the burst-overlap regions. Unlike interseismic or postseismic scenarios, where sparse coherent points require stable-scatterer selection and SHP-based phase linking to enhance the SNR, coseismic deformation benefits from high interferometric coherence due to short temporal baselines (e.g., days to weeks) and strong displacement signals that dominate noise. The MTSB method leverages these conditions through its block-wise periodogram approach, which directly estimates misregistration using burst-overlap interferograms, eliminating the need for stable-scatterer selection or advanced phase enhancement. This streamlined workflow significantly simplifies processing while achieving higher accuracy than pixel-wise displacement estimation, as demonstrated in Fig. 3.

The 2015 $M_w$ 8.3 Illapel earthquake ruptured the central Chile subduction zone, where the Nazca Plate subducts beneath the South American Plate along the Peru-Chile trench (Figure 9). The earthquake occurred offshore near the Illapel-Coquimbo region and generated large coastal deformation, making it a well-observed megathrust event for testing geodetic measurements of coseismic displacement. This event has previously been analyzed using Sentinel-1 TOPS along-track and across-track interferometry by *Grandin et al.* (2016), who demonstrated that burst-overlap interferometry can recover the large along-track coseismic displacement and, when combined with conventional across-track InSAR, resolve the three-dimensional coseismic displacement field with subdecimetric agreement with GNSS observations.

Here, we use this well-documented event as a benchmark to test whether MTSB can recover the known coseismic along-track signal within a multi-track, ITRF2014-referenced processing framework. Using the CtSent toolbox, we processed pre- and post-earthquake Sentinel-1 SAR image pairs from ascending (20150826-20150919) and descending (20150824-20150917) tracks. A 2.5 km × 2.5 km block-wise periodogram is then directly applied to the burst-overlap interferograms to derive the coseismic displacement field. The resulting displacement field, which encompasses both plate motion and coseismic displacement components, is referenced to the ITRF2014 frame.

To evaluate the performance of MTSB, we isolated earthquake-induced deformation from ITRF2014 displacement by calculating and subtracting secular plate motion at each grid point (Fig. S9), though the contribution from plate motion is minor relative to coseismic deformation. For ground truth validation, GNSS data from the Nevada Geodetic Laboratory *(Blewitt et al., 2018)* were projected into the along-track direction for comparison (Fig. 9). The MAE between GNSS and MTSB estimates is 2.78 cm (ascending) and 3.59 cm (descending), representing improvements of approximately 10% and 9% for the ascending and descending tracks, respectively, relative to the conventional BOI baseline (Fig. S10). These results demonstrate the applicability of MTSB to this favorable megathrust coseismic case. For near-field strike-slip earthquakes, however, the strain model should not be imposed continuously across a mapped fault rupture, where a true displacement discontinuity may occur. In such settings, rupture masking, separate treatment of the two fault blocks, smaller adaptive blocks near the rupture, or discontinuity-aware regularization may be required to preserve the coseismic displacement step.

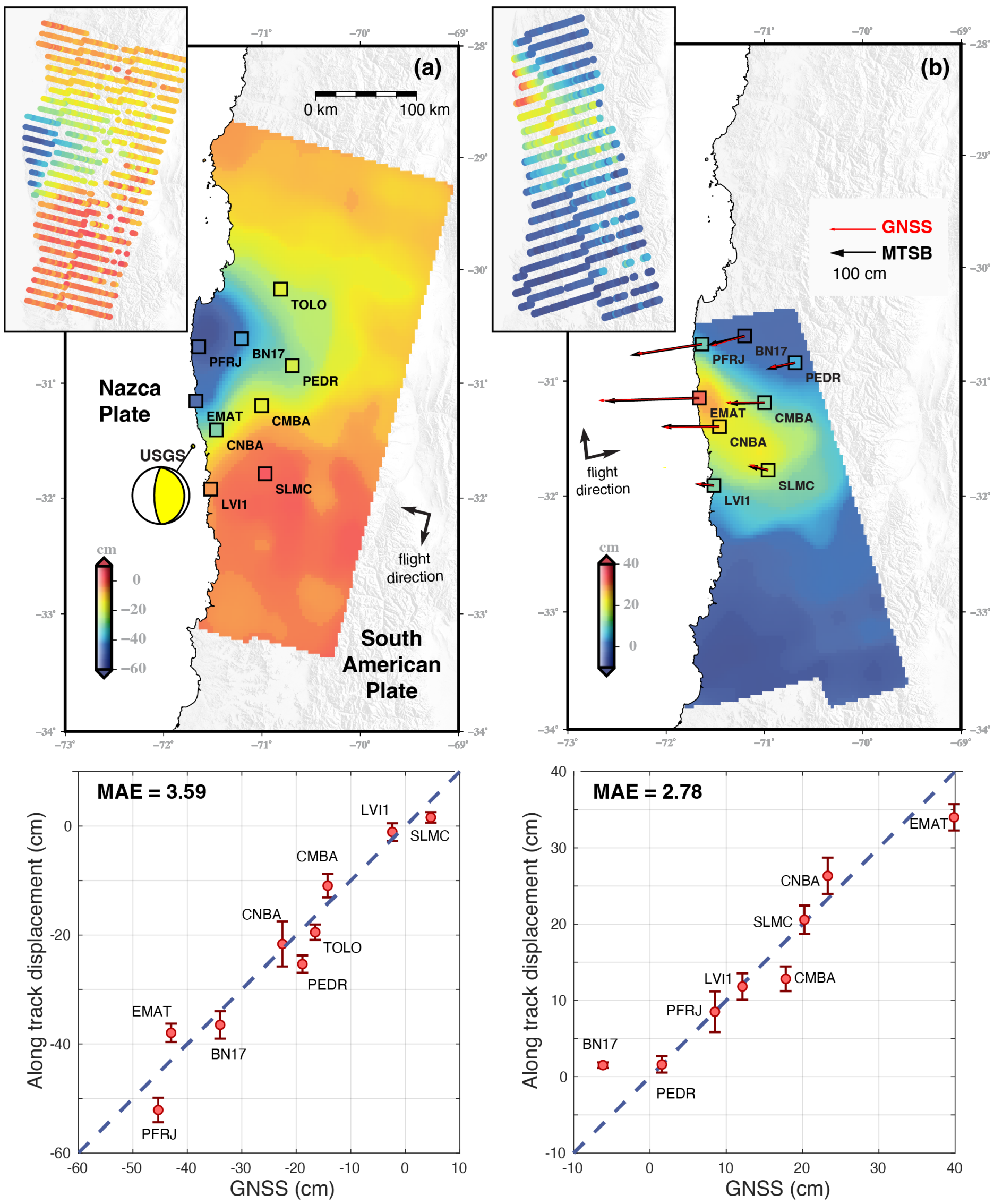

**Figure 9.** Coseismic along-track displacement fields (inset) and interpolation for descending (a) and ascending (b) tracks, as measured by the MTSB method. The lower panels present cross-plots comparing GNSS and MTSB data with error bars indicating uncertainties of the MTSB estimates. The coseismic along-track displacement pattern is broadly consistent with the Sentinel-1 TOPS burst-overlap results reported by *Grandin et al.* (2016).

In this favorable offshore megathrust case, the block-wise periodogram improves local phase estimation while preserving the smoothly varying coseismic signal sampled by the burst-overlap regions. Decomposition of the displacement field into east–west and north–south components revealed small residuals relative to the GNSS observations (Fig. 9b). The arrows in Fig. 9b represent the projected east–west and north–south displacement components reconstructed from the decomposed MTSB measurements. These results show that MTSB preserves the large-amplitude, smoothly varying coseismic signal sampled by the burst-overlap regions and yields along-track displacements consistent with previous BOI results and independent GNSS observations.

## 5 Discussion

### 5.1 Advances of MTSB beyond conventional BOI

Conventional BOI has demonstrated the capability of recovering along-track deformation, but its broader application has remained limited by orbit-dependent reference frames, residual azimuth misregistration, fading-related phase degradation, limited consistency among independent tracks, and the lack of a unified multi-track time-series framework capable of preserving long-wavelength tectonic signals. MTSB addresses these limitations by preserving long-wavelength tectonic deformation, correcting orbit-dependent biases through time-series inversion, and expressing observations from multiple Sentinel-1 tracks within a common ITRF2014 reference frame. This unified framework extends conventional BOI from localized, track-specific measurements toward spatially continuous, geodetically referenced deformation fields that can be directly integrated with GNSS and incorporated into regional geodynamic analyses.

The interseismic results along the southern Dead Sea Fault provide the clearest demonstration of this advance. *Li et al.* (2021) showed that burst-overlap observations can resolve millimeter-per-year fault-parallel motion in this region. MTSB preserves this capability while substantially improving the spatial continuity and internal consistency of the velocity field across adjacent Sentinel-1 tracks relative to previous BOI processing. By reducing track-to-track discontinuities and preserving the long-wavelength ITRF2014-referenced velocity field, the resulting velocity field provides a more stable basis for estimating the fault-parallel velocity gradient, yielding a slip rate of 4.8 mm/yr and a locking depth of 8.5 km. These results strengthen the reliability of the inferred fault kinematics and shallow locking behavior, while remaining consistent with previous BOI and GNSS interpretations.

The Türkiye earthquake sequence further demonstrates the capability of MTSB to recover spatially dense and temporally coherent postseismic deformation over a broad region. The recovered along-track time series agree well with projected GNSS observations, indicating that MTSB captures the first-order temporal evolution of the postseismic response while providing a spatially continuous displacement field that more completely resolves postseismic deformation across the fault system during the first year after the earthquake. The MTSB results therefore provide complementary along-track observations that can be integrated with LOS InSAR and three-component GNSS data in future studies to test competing postseismic mechanisms.

The Illapel earthquake provides an independent benchmark for evaluating MTSB in a large coseismic deformation setting. *Grandin et al.* (2016) demonstrated that Sentinel-1 BOI can recover the coseismic along-track displacement associated with this event. We therefore use the Illapel case not to provide a new rupture interpretation, but to evaluate whether MTSB preserves the previously established large-amplitude coseismic signal after multi-track referencing and residual-misregistration correction. Geophysically, the resulting spatially continuous along-track field provides an additional along-track constraint on the lateral variation of coseismic displacement and can strengthen three-dimensional displacement and slip reconstructions when combined with LOS InSAR and GNSS observations.

Taken together, the three case studies show that the MTSB workflow can be applied across deformation regimes that differ substantially in magnitude, temporal behavior, and spatial scale. Quantitative validation metrics for the case-specific results are summarized in Table S6, including internal multi-track consistency, GNSS-based RMSE/MAE/bias, and residual uncertainty estimates. These comparisons place MTSB in the context of previous BOI studies and clarify that its main contribution is a unified, ITRF2014-referenced, multi-track time-series framework for BOI-based deformation measurements, with explicit GNSS validation and uncertainty assessment.

### 5.2 Postseismic signal characteristics from MTSB measurements

The Türkiye case demonstrates that MTSB can recover spatially continuous and temporally coherent postseismic deformation at the regional scale while maintaining consistency across ascending and descending tracks. Agreement with projected GNSS time series supports interpreting the recovered signals as postseismic deformation rather than residual misregistration. The resulting dense along-track measurements provide spatial information complementary to InSAR and GNSS observations. Because BOI is primarily sensitive to the approximately north–south along-track component of horizontal motion, MTSB preferentially resolves deformation with a strong north–south component but is less sensitive to predominantly east–west motion. This directional limitation is particularly relevant to the Türkiye earthquake doublet because of its multiple ruptured segments with different strike orientations.

The observed deformation may reflect contributions from afterslip, viscoelastic relaxation, poroelastic rebound, or a combination of these processes. Because these mechanisms can produce partially overlapping surface deformation patterns, especially when only one displacement component is considered, the MTSB results should be viewed as an important complementary constraint rather than a standalone basis for rheological interpretation. In particular, MTSB provides dense along-track horizontal displacement measurements that complement conventional LOS InSAR and sparse three-component GNSS observations, thereby improving constraints on the spatial and temporal evolution of postseismic deformation.

### 5.3 MTSB-based Euler pole determination

For continental-scale studies, conventional InSAR suffers from long-wavelength noise and the fundamental constraint of providing only relative, rather than absolute, displacements. By contrast, the MTSB framework preserves absolute, long-wavelength tectonic motion referenced to the ITRF2014 frame, while simultaneously removing orbital and other systematic errors. To estimate the Euler pole of the Arabian plate, we employed a random-walk Metropolis-Hastings MCMC scheme separately to the GNSS-only and MTSB-only data sets and jointly to the combined data set. The algorithm explores the parameter space of longitude, latitude, and angular velocity of the Euler pole, with physically reasonable upper and lower bounds. Specifically, independent uniform priors were assigned to the Euler-pole longitude, latitude, and angular velocity over the ranges 0°–360°, 40°–60°, and 0.4°–0.6°/Ma, respectively. The chain was initialized at 340°, 50°, 0.4°/Ma. At each iteration, the three parameters were perturbed independently using uniform random-walk proposals with step scales of 0.5°, 0.5°, 0.01°/Ma. The resulting proposal covariance matrix was diagonal, and proposals crossing the prior bounds were reflected back into the permitted parameter ranges. Posterior parameter covariance was not prescribed but was estimated empirically from the retained post-burn-in samples. Candidate models were evaluated using an empirical pseudo-likelihood based on the mean absolute residual between the observed and predicted along-track velocities and were accepted or rejected according to the Metropolis–Hastings criterion.

We performed 100,000 iterations and discarded the first 20,000 samples as burn-in. Parameter summaries, covariance estimates, and sampling diagnostics were calculated from the remaining 80,000 samples. Sampling convergence was evaluated using trace plots, autocorrelation functions, effective sample sizes (ESS), and posterior parameter correlations (Supplementary Figures S11–S14). The Euler-pole inversion included both near-field and far-field MTSB observations. Using the full velocity field increased spatial coverage and improved constraints on the regional rigid-plate rotation. Although localized interseismic deformation near the Dead Sea Fault contributes to the residual field, the inversion is primarily controlled by the long-wavelength Arabian plate-motion signal because of its broader spatial extent and larger amplitude.

Our MCMC-based joint Euler pole solution in Fig. 10 using MTSB observations and GNSS data is Lon = 354.97°, Lat = 50.68°, ω = 0.529°/Ma (ITRF2014), which is broadly consistent with recent determinations. Relative to published models, our solution is close to GNSS-based estimate of *Viltres et al.* (2022) (Lon = 353.91°, Lat = 50.93°, ω = 0.524°/Ma) and *Liu et al.* (2025) (Lon ≈ 351-352°, Lat ≈ 49-50°, ω ≈ 0.552°/Ma). Compared with the GNSS solution of Viltres et al. (2022), our result exhibits nearly identical longitude, a slightly northerly latitude, and a somewhat higher angular velocity, but overall falls within the range of published models. By contrast, the InSAR-derived solution of *Liu et al.* (2025) yields a pole shifted to the southwest and consistently higher angular velocity relative to GNSS. The results indicate that, compared to conventional InSAR, our approach removes the orbital errors and provides direct sensitivity to along-track motion, which is dominated by the north–south component. To evaluate the spatial fit of the Euler-pole model, we compared the observed MTSB along-track velocity field with the velocity predicted by the best-fitting Euler pole and calculated the corresponding residual velocity field (Fig. S15). The best-fitting Euler model reproduces the first-order regional velocity gradient, while the residual field represents localized departures from the rigid-plate model arising from deformation, measurement uncertainty, and other unmodeled processes. The residual field is centered close to zero, with a mean of −0.046 cm/yr, a median of −0.002 cm/yr, and an RMSE of 0.30 cm/yr. The median absolute residual is 0.17 cm/yr, and 70% of the observations have absolute residuals smaller than 0.25 cm/yr.

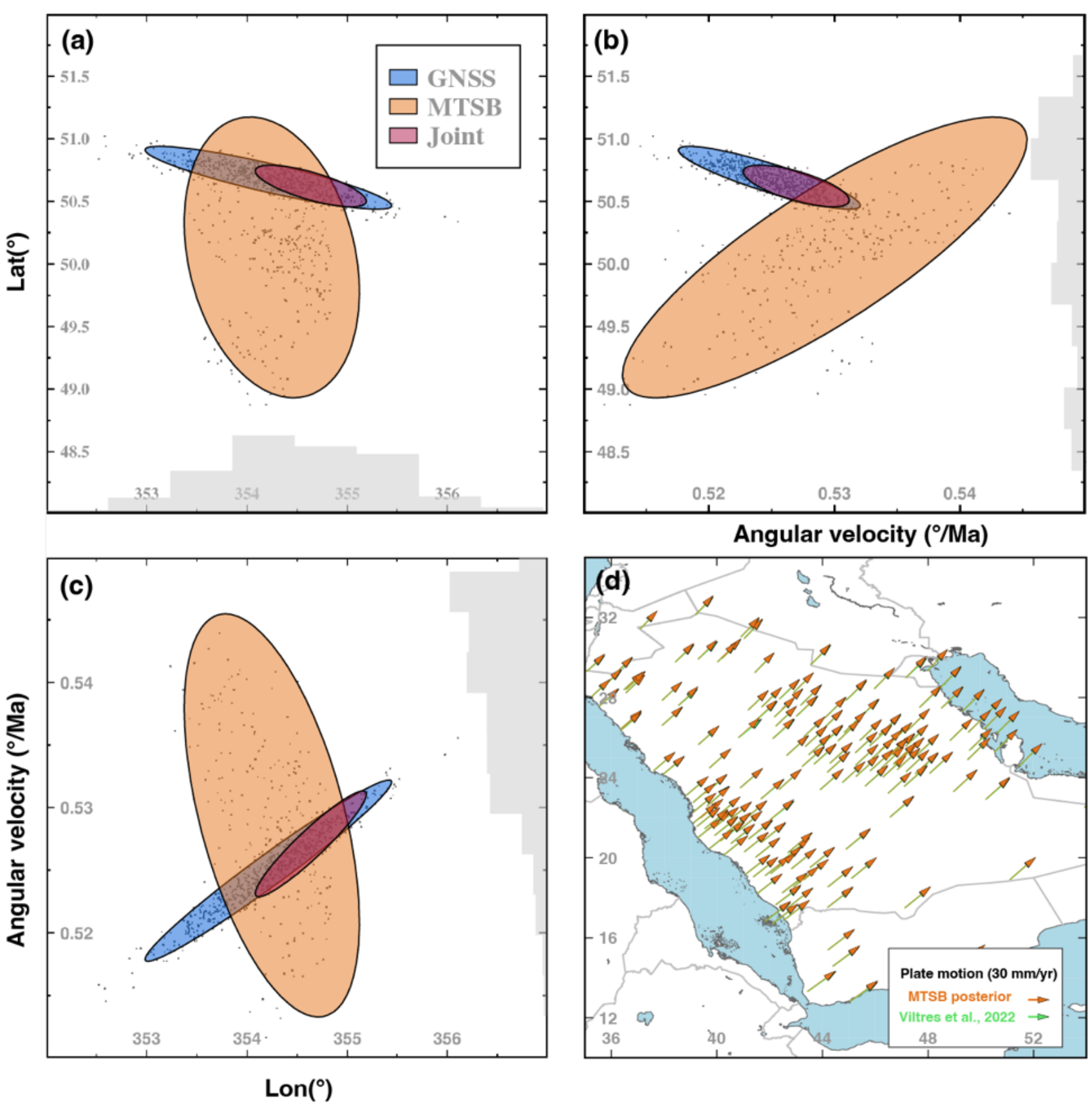

**Figure 10.** (a–c) Arabian plate Euler pole solutions and uncertainties estimated from MTSB and GNSS velocities in the ITRF2014 reference frame. Marginal posterior distributions of the three Euler pole parameters are shown as histograms. The point clouds represent Monte Carlo posterior samples (blue: GNSS, orange: MTSB, purple: joint solution). The shaded ellipses denote 95% covariance ellipses calculated from the posterior sample means and covariance matrices. (d) Predicted plate-motion vectors from the Euler pole estimated using MTSB velocities compared with published GNSS velocities *(Viltres et al., 2022)*.

### 5.4 Orbital errors and improvements in image coregistration

The MTSB algorithm first eliminates the reliance on ESD in the image coregistration process, which traditionally suppresses large-scale deformation signals such as plate motion and rapid fault slip, as demonstrated across all case studies. The residual orbital errors induced a constant misregistration shift into the BOI interferograms, as demonstrated in Fig. 5. The key advancement of MTSB is the explicit correction of orbit-induced misregistration errors through a time-series inversion framework, which models and separates orbital artifacts from the deformation signal. The correction improves the performance of the image coregistration. As shown in Fig. 11, the constant misregistration shift is corrected through the time-series inversion framework. This step ensures that displacements from multiple orbits are consistently referenced within a unified geodetic frame, i.e., ITRF2014, which significantly improves cross-track continuity and suppresses systematic biases that often compromise multi-track analyses. The effectiveness of this approach is clearly demonstrated in the interseismic case study along the Dead Sea Fault, where MTSB resolves a spatially continuous velocity field with a slip rate of 4.8 mm/yr and a locking depth of 8.5 km, both of which are consistent with geological and GNSS-derived benchmarks.

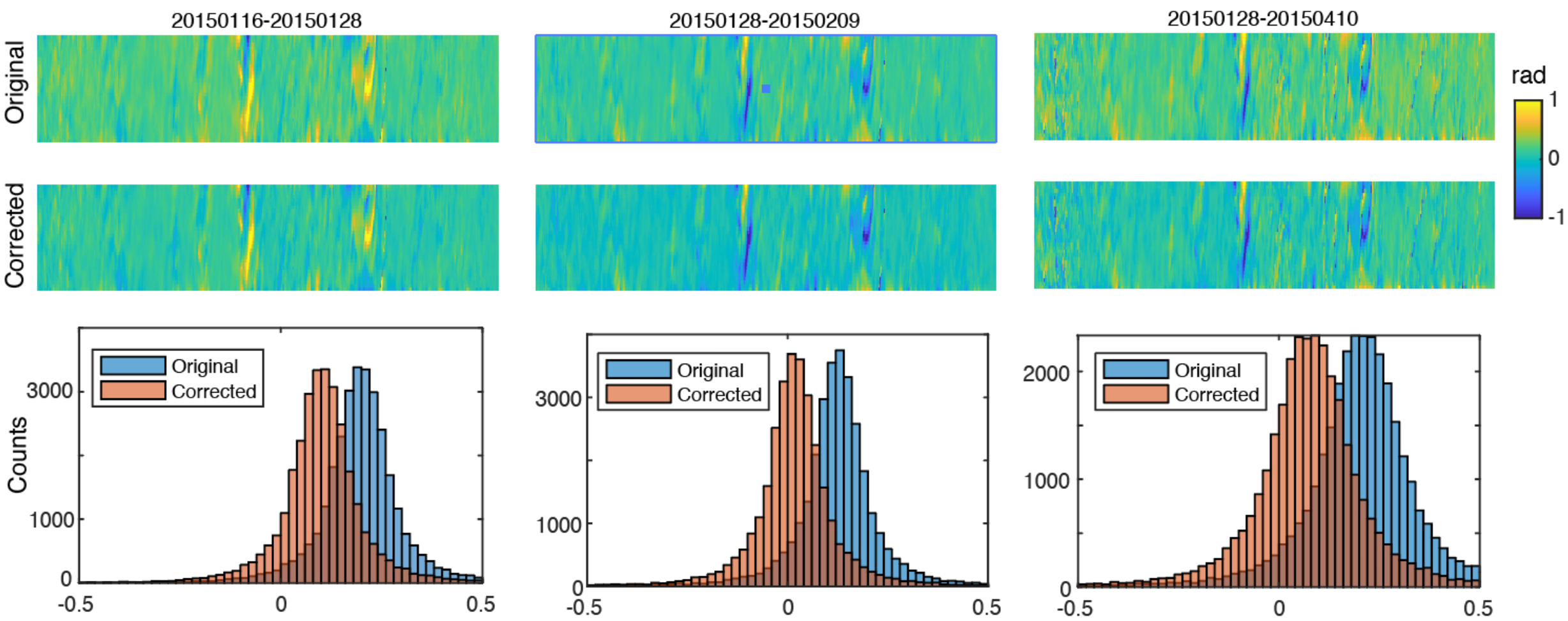

**Figure 11.** Top row: examples of original unwrapped burst-overlap interferograms. Middle row: corresponding interferograms after orbital-error correction. Bottom row: histograms illustrating the orbital error shift between the original and corrected interferograms.

### 5.5 Importance of fading signal correction

Fading signal has recently been recognized as an important issue in InSAR time-series analysis, as it can significantly degrade phase quality and bias the estimation of surface deformation *(Ansari et al., 2021; Ma  et al., 2025; Maghsoudi et al., 2022)*. In the context of BOI, horizontal deformation is exploited by double differencing between forward- and backward-looking interferograms. While this differencing step effectively reduces common-mode errors such as tropospheric delay, the differing look geometries of the forward and backward interferograms make BOI particularly susceptible to fading signal. Specifically, fading manifests as localized loss of coherence or signal dropouts in one viewing geometry but not the other, leading to spurious discontinuities in the differenced BOI interferograms. Such artifacts may obscure the true azimuthal deformation signal and complicate subsequent phase unwrapping and time-series inversion. As demonstrated in Fig. 12, our simulation experiments clearly show how fading signal can introduce systematic artifacts in BOI interferograms, even when other error sources are well controlled. These results highlight the necessity of explicitly accounting for fading signal effects in BOI processing to ensure robust estimation of along-track deformation and reliable long-term monitoring.

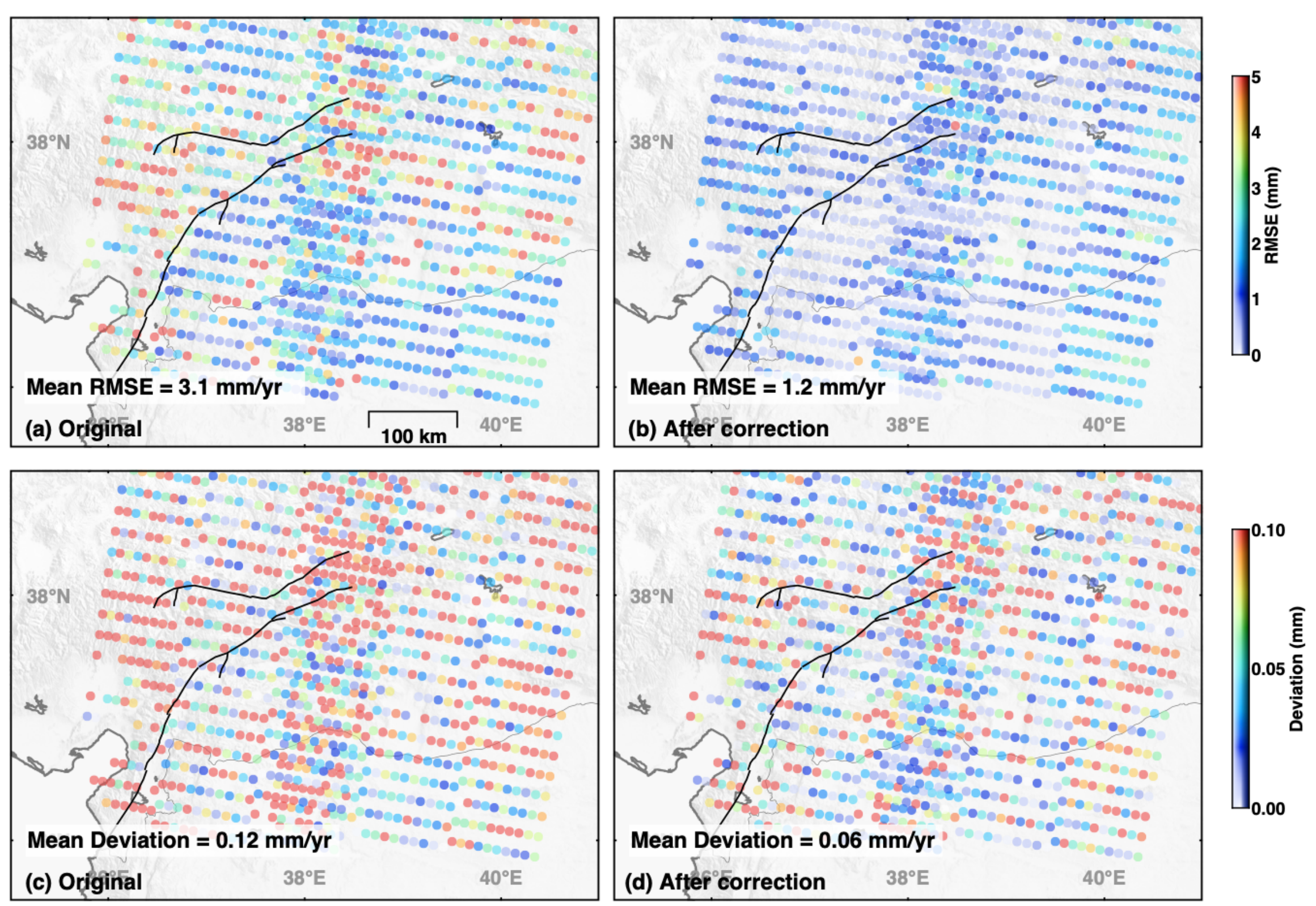

**Figure 12.** Root-mean-square error (RMSE) and deviation maps of the time-series results before and after fading-signal correction. Panels (a) and (c) show the RMSE and the absolute mean residual (deviation) of the original solution, respectively, while panels (b) and (d) present the corresponding results after applying the fading-signal correction.

### 5.6 Challenges and Prospects

Despite the advances of MTSB, the accuracy and precision of MTSB measurements remain dependent on deformation magnitude, temporal behavior, and coherence conditions. Case-specific formal uncertainties, empirical consistency measures, and external GNSS validation results are reported in Sections 4.1-4.3 and summarized in Table S6. Together, these assessments support millimeter-per-year accuracy for interseismic velocities and centimeter-level accuracy for postseismic and coseismic displacements. We further evaluated the coherence dependence of MTSB uncertainty using the Türkiye postseismic dataset, which spans a broad range of coherence conditions (Table S8). Because the postseismic deformation is transient and nonlinear, uncertainty was quantified for cumulative displacement rather than for a constant velocity. For each MTSB block, the formal 1σ uncertainty of cumulative displacement was calculated from the covariance of the displacement difference between the final and reference epochs, with the first epoch fixed as the displacement reference. The blocks were grouped into four mean-coherence bins: $\gamma > 0.7$, $0.5 < \gamma \leq 0.7$, $0.3 < \gamma \leq 0.5$, and $\gamma \leq 0.3$. For track T123, the median uncertainties across the three interferometric wide swaths increase from 0.21-0.33 cm for $\gamma > 0.7$ to 0.71-1.22 cm for $0.3 < \gamma \leq 0.5$. For track T21, the corresponding uncertainties increase from 0.22-0.58 cm to 1.08-1.27 cm. The lowest-coherence bin contains only three valid blocks for T123 and none for T21, limiting the robustness of its uncertainty statistics. These results demonstrate that reduced coherence generally increases the uncertainty of the recovered cumulative displacement.

The computational cost of MTSB is mainly controlled by the Seq + EMI phase-linking step. For the Türkiye case, which includes 24 Sentinel-1 acquisitions and 54 burst-overlap regions across descending tracks, this step required approximately 6 min per burst-overlap region on a Linux workstation with 64 physical cores, 128 logical threads, and 502 GiB RAM using a 20-worker MATLAB parallel pool. If processed sequentially, all 54 burst-overlap regions would require approximately 5.4 h. Because individual burst-overlap regions can be processed independently, the workflow is naturally parallelizable, and the total wall-clock time depends on the number of available workers and resource allocation. Other MTSB-specific steps, including fading-signal correction, spatial smoothing, and time-series inversion, were substantially faster and did not dominate the total runtime. These timings exclude standard SAR preprocessing.

The current MTSB workflow does not explicitly correct for solid Earth tides, ocean tidal loading, or ionospheric disturbances. Although solid Earth tides and ocean tidal loading are important corrections in conventional InSAR processing, they are expected to be strongly attenuated through BOI double differencing because the forward- and backward-looking burst observations are acquired within the same satellite pass. Small residual effects may nevertheless remain because of incomplete cancellation or residual spatial gradients. In contrast, BOI is more susceptible to ionospheric disturbances, particularly long-wavelength azimuthal TEC gradients, which can introduce apparent azimuth shifts and long-wavelength interferometric phase ramps.

The sensitivity of MTSB to residual ionospheric effects depends on both the magnitude and spatial structure of the azimuthal ionospheric phase gradients relative to the tectonic signal. Temporally intermittent ionospheric disturbances may be partially suppressed through temporal low-pass filtering, multi-epoch time-series inversion, and averaging across multiple burst-overlap observations. Consequently, their influence on long-term deformation rates is expected to be limited in many cases. In contrast, ionospheric disturbances that produce systematic long-wavelength gradients are more problematic because their signals may be partly absorbed into the estimated long-wavelength velocity field or into the residual misregistration. To quantify the magnitude of the remaining long-wavelength artifacts, including possible ionospheric contributions, we fitted a planar ramp to the corrected burst-overlap residual at each epoch and used the peak-to-peak amplitude of the fitted ramp as an empirical measure of these residual signals. These ramps are not unique indicators of ionospheric contamination but provide a useful measure of its possible contribution. For the Dead Sea T123 track, the median residual ramp amplitudes of individual burst-overlap regions range from 0.45 to 1.07 cm (Table S9 and Fig. S16). Across all three tracks, the largest temporal-median amplitude is 1.30 cm for T94 Ovl7. Despite these centimeter-level epoch-wise residuals, the temporal evolution of the fitted ramps exhibits only weak systematic trends. For each overlap region, we estimated the long-term trend of the residual-ramp time series using linear regression. The resulting absolute slopes, interpreted as equivalent velocity contributions, are below 1 mm/yr for most overlap regions. These results suggest that the remaining long-wavelength artifacts, including possible ionospheric contributions, are unlikely to dominate the recovered first-order interseismic velocity field for the analyzed Dead Sea dataset and observation period.

Further mitigation of residual non-tectonic signals remains desirable, particularly for applications requiring higher accuracy or conducted under severe ionospheric conditions. Although solid Earth tides and ocean tidal loading are expected to be strongly attenuated through BOI double differencing, acquisition-specific corrections based on established geophysical models could be incorporated to account for incomplete cancellation and residual spatial gradients *(DiCaprio and Simons, 2008; Yu et al., 2020)*. Split-spectrum techniques may be used to estimate dispersive ionospheric phase contributions directly from SAR observations *(Gomba et al., 2016)*. Time-series approaches that jointly estimate acquisition-wise ionospheric screens from multiple interferograms may further improve the correction of long-wavelength ionospheric artifacts in conventional InSAR and BOI measurements *(Piromthong et al., 2026)*. GNSS-TEC products can provide complementary external constraints on regional-scale ionospheric delays and gradients, although their spatial resolution may be insufficient to characterize fine-scale ionospheric irregularities *(Wang et al., 2026)*. Combining these approaches with residual-based quality control would improve the robustness of MTSB under more challenging observing conditions.

## 6 Conclusions

The MTSB algorithm addresses critical limitations of conventional BOI by enabling the retrieval of large-scale along-track horizontal deformations across tectonic boundaries. Unlike traditional ESD techniques, which suppress tectonic signals (e.g., plate motion, rapid fault slip) and introduce systematic biases in multi-track configurations, MTSB overcomes these challenges through a three-stage framework incorporating a unified time-series framework that separates deformation and residual misregistration. Independent comparisons with GNSS indicate centimeter-level agreement for postseismic and coseismic displacements and millimeter-per-year agreement for interseismic velocities. By preserving absolute plate-motion information, MTSB can be directly integrated with GNSS velocities to improve constraints on Euler pole and plate rotation estimates. The resulting spatially dense deformation fields also provide complementary observations for future investigations of postseismic processes. Overall, MTSB represents a methodological advance in SAR-based horizontal deformation retrieval by bridging high-spatial-resolution along-track measurements with the ITRF2014 reference frame. This capability has potential for quantifying plate-boundary strain accumulation, supporting future investigations of lithospheric rheology, and advancing regional to continental seismic hazard assessment.

**Acknowledgments**

This work was supported by the University of Texas at Austin under grant 201503664. The figures were generated using MATLAB (R2023b, The MathWorks, Natick, MA, USA) and Generic Mapping Tools 6.1.0 software *(Wessel et al., 2013)*.

**Data Availability Statement**

The raw Sentinel-1 data were downloaded from the Alaska Satellite Facility (https://vertex.daac.asf.alaska.edu/). The CtSent software used for Sentinel-1 preprocessing is available at https://zenodo.org/records/10776079. The principal CtSent and MTSB processing parameters used in this study are provided in Tables S2 and S3. The GNSS time series data are available from the Nevada Geodetic Laboratory *(Blewitt et al., 2018)*.

**Conflict of Interest Disclosure**

The authors declare there are no conflicts of interest for this manuscript.

**Author Contributions**

**X. Li:** Conceptualization, Methodology, Formal analysis, Writing – original draft, Visualization

**Z. Gao:** Visualization, Review & editing

**H. Chen:** Visualization, Review & editing.

**Y.K. Chen:** Review & editing.

**Z.F. Ma:** Conceptualization, Methodology, Visualization, Review & editing.

**A. Savvaidis:** Review & editing.